\documentclass[prd,nofootinbib,english,a4paper]{revtex4-2}
\usepackage[utf8]{inputenc}
\usepackage{amsmath}
\usepackage{amsfonts}
\usepackage{amssymb}

\usepackage{xcolor}
\usepackage{accents}
\usepackage{mathrsfs}
\usepackage{hyperref}
\hypersetup{
	colorlinks=true,
	linkcolor=blue,
	filecolor=magenta,
	citecolor=blue
}
\usepackage{graphicx}
\usepackage{tikz}
\usetikzlibrary{external,decorations.pathmorphing,fpu}
\allowdisplaybreaks

\usepackage{amsmath}

\usepackage{bookmark}

\begin{document}

    \title{Symmetric Teleparallel Connection and Spherical Solutions in Newer GR}

	\author{Manuel Hohmann}
	\email{manuel.hohmann@ut.ee}
	\affiliation{Laboratory of Theoretical Physics, Institute of Physics, University of Tartu, W. Ostwaldi 1, 50411 Tartu, Estonia}

	\author{Vasiliki Karanasou}
	\email{vasiliki.karanasou@ut.ee}
	\affiliation{Laboratory of Theoretical Physics, Institute of Physics, University of Tartu, W. Ostwaldi 1, 50411 Tartu, Estonia}

\begin{abstract}
\begin{center}
\textbf{Abstract}
\end{center}
In this article, we focus on symmetric teleparallel gravity, a modification of General Relativity where gravity is described by the non-metricity of an affine connection, whose curvature and torsion vanish. In these theories, the fundamental variables are the metric and an affine connection. Starting from the coincident gauge, a system of coordinates for which the affine connection coefficients vanish, we derive the most general connection for a spherically symmetric and stationary spacetime. We then derive the field equations in a specific class of symmetric teleparallel theories, the so-called Newer General Relativity. This theory is constructed from the five possible quadratic scalars of non-metricity. We find two families of vacuum solutions that correspond to some exotic objects and we study their properties. In particular, we investigate the possibility of having a traversable wormhole, we compute the Komar mass, we discuss the conditions for asymptotic flatness, the existence of singularities, the radial motion and bound orbits of particles around these objects, the light deflection as well as the causal structure of these spacetimes.
\end{abstract}
\maketitle
\section{Introduction}
General Relativity (GR) has been a successful theory of gravity considering that a great variety of problems in astrophysics and cosmology have been addressed in this context and several experimental tests have verified its validity  \cite{10.1119/1.1935800, doi:10.1126/science.129.3349.621, PhysRevD.66.082001,Fomalont:2009zg, PhysRevLett.13.789, PhysRev.170.1186, Williams:2004qba, Baker:2014zba, Will:2014kxa, Will:1993hxu,Turyshev:2008ur, Asmodelle:2017sxn, https://doi.org/10.1002/andp.19163540702}. However, it appears that this theory is not the final answer since it is not able to explain the observations related with the accelerated expansion of the universe \cite{SupernovaCosmologyProject:1998vns,SupernovaSearchTeam:1998fmf} as well as several cosmological tensions that arise \cite{DiValentino:2021izs,Schoneberg:2021qvd,Abdalla:2022yfr}. The standard cosmological model explains the expansion of the universe by assuming the existence of dark energy in the context of GR \cite{Copeland:2006wr,Planck:2018vyg}. However, there is little information about the properties of this entity and no experimental data that validates its existence. At the same time, GR is not compatible with the principles of quantum field theory. Modified theories of gravity are proposed as an alternative in order to address these issues.

There are several ways to modify the gravitational theory. In GR, gravity is the result of the curvature of the spacetime, which depends on the Levi-Civita connection. A simple generalization of GR, the $f(R)$ theory of gravity, arises by substituting the Ricci scalar $R$ in the gravitational action with an arbitrary function $f(R)$ \cite{Capozziello:2011et,Sotiriou:2008rp}. Assuming a general affine connection, we also introduce the quantities torsion and non-metricity along with the curvature~\cite{Hehl:1994ue,JimenezCano:2021rlu, BeltranJimenez:2019odq, BeltranJimenez:2019esp}. A particularly interesting class of these theories are the teleparallel theories of gravity, where the gravitational field is mediated by either the torsion (metric teleparallel theories) or non-metricity (symmetric teleparallel theories) while the curvature vanishes \cite{Hohmann:2022mlc}.

In this paper, we focus on symmetric teleparallel theories, where the curvature and torsion vanish and only non-metricity is present \cite{Nester:1998mp, Adak:2005cd, Mol:2014ooa, Adak:2018vzk}. The fundamental variables of these theories are the metric and the affine connection. The simplest symmetric teleparallel theory of gravity is the so-called Symmetric Teleparallel Equivalent of GR (STEGR). This theory is described by an action similar to the Einstein-Hilbert action where the Ricci scalar is substituted by the non-metricity scalar and it is equivalent to GR up to a boundary term in the action, which does not contribute to the field equations. A simple generalization of STEGR that we will consider in this work, is the Newer General Relativity \cite{BeltranJimenez:2017tkd}. In this theory, the non-metricity scalar is decomposed into its five irreducible components and five parameters are introduced in the Lagrangian. Different values of these parameters describe different theories of gravity. In fact, it has been proved that in order to reproduce the post-Newtonian limit, the parameters should satisfy some conditions \cite{Flathmann:2020zyj} that can be classified into two different families. We shall call Type 1 Newer GR the theory that corresponds to the first family and Type 2 Newer GR the theory that corresponds to the second family enhanced with the ghost-free condition \cite{Bello-Morales:2024vqk}. Both types are 1-parameter theories. It is worthy to mention the similarity with the New General Relativity, a metric teleparallel theory of gravity which is constructed by the irreducible components of the torsion scalar. This theory is more explored in the literature \cite{Hayashi:1979qx, BeltranJimenez:2019nns, Blixt:2018znp,Hohmann:2019sys, Blixt:2019ene,Guzman:2020kgh,  Ualikhanova:2019ygl, Hohmann:2018jso, Golovnev:2023uqb, Asukula:2023akj, Bahamonde:2021gfp}.

Various topics in Newer GR have been already investigated, in particular the parameterized post-Newtonian limit \cite{Flathmann:2020zyj}, the polarization of gravitational waves \cite{Hohmann:2018xnb} as well as their propagation \cite{Hohmann:2018wxu} and the Hamiltonian analysis \cite{DAmbrosio:2020nqu}. Several other  modifications have been studied. $f(Q)$ gravity is the analogue of $f(R)$ in the symmetric teleparallel context. However, despite the fact that STEGR is equivalent to GR, $f(Q)$ is not equivalent to $f(R)$ gravity. Other examples are the $f(Q,B)$ gravity, where $B$ is a boundary term between the Ricci scalar and the non-metricity scalar \cite{Capozziello:2023vne}, scalar non-metricity theories \cite{Runkla:2018xrv, Jarv:2018bgs} and the Gauss-Bonnet teleparallel theories \cite{Armaleo:2023rhj}. In order to gain a better understanding of the gravitational theory we are interested in, as well as its viability, and to obtain more information about black holes and other compact objects, it is useful to study the spherical solutions of the theory. Spherical solutions have been investigated in various symmetric teleparallel extensions \cite{Adak:2004uh, Adak:2008gd, Lin:2021uqa}. Two sets of components that correspond to the symmetric teleparallel connection for a stationary and spherically symmetric geometry have been found in~\cite{DAmbrosio:2021zpm}, and spherical solutions in $f(Q)$ gravity are studied.

In symmetric teleparallel gravity, it is usually convenient to work in the coincident gauge. This is a system of coordinates for which the affine connection vanishes identically \cite{Koivisto:2019ggr, DAmbrosio:2020nqu, Bahamonde:2022zgj}. In our work, starting from this gauge, we derive the components of the affine connection for any spherically symmetric and stationary spacetime. We realize that these components correspond to the second set of \cite{DAmbrosio:2021zpm} and we show that the first set is a subcase of the first one without taking any limits. Thus, our result includes both sets. We then focus on Newer General Relativity and we derive the field equations. Since solving the field equations for the most general class of theories is a rather intricate task, we restrict ourselves to specific and physically motivated classes of theories. For these we find one family of solutions in Type 1 Newer GR and a common solution for both Type 1 and Type 2 theories. These solutions correspond to some exotic objects that are not predicted  in GR. We investigate their characteristics such as their Komar mass, the conditions for asymptotic flatness, the existence of singularities and horizons, the radial motion and bound orbits of particles near these objects, the light deflection and their causal structure.

The paper is organized in the following way: In Section \ref{sec:symmtel}, the fundamentals of symmetric teleparallel gravity are introduced and we derive the most general connection for any stationary and spherically symmetric spacetime by using the coincident gauge. In Section \ref{sec:ngr}, we discuss a specific class of symmetric teleparallel gravity, the Newer General Relativity, we discuss the difference between Type 1 and Type 2 theories and we derive the field equations in this theory. In Section \ref{sec:solutions}, we present two interesting families of solutions, one for the Type 1 theory and one for both Type 1 and Type 2 theories and we study its properties. In particular, we investigate if they could describe traversable wormholes, we calculate the Komar mass and the gravitational redshift, we study the conditions for asymptotic flatness, the existence of singularities, the motion of particles near these objects and their causal structure. We summarize the conclusions of this work in Section \ref{sec:conclusions}. For completeness, we display the spherically symmetric field equations of Type 1 Newer GR in appendix~\ref{appendix1} and of Type 2 Newer GR in appendix~\ref{appendix2}.

The notations we adopt in this article are the following: Greek indices label coordinates of the spacetime and they run from $0$ to $3$. The Minkowski metric components are denoted by $\eta_{\mu\nu} = \textrm{diag}(-1,1,1,1)$. A ring above quantities like $\mathring{R}$ describes quantities constructed with the Levi-Civita connection.

\section{Symmetric teleparallel connection}
\label{sec:symmtel}
In symmetric teleparallel gravity, the dynamical variables are the metric $g_{\mu\nu}$ and the affine connection $\Gamma^{\rho}_{\;\mu\nu}$. The gravitational field is mediated by the non-metricity while the curvature and the torsion vanish identically \cite{Nester:1998mp, Adak:2005cd, Mol:2014ooa, Adak:2018vzk, Hohmann:2022mlc}. This implies
\begin{subequations}
\begin{align}
R^{\rho}_{\;\sigma\mu\nu}&=\partial_{\mu}\Gamma^{\rho}_{\;\sigma\nu}-\partial_{\nu}\Gamma^{\rho}_{\;\sigma\mu}+\Gamma^{\rho}_{\;\lambda\mu}\Gamma^{\lambda}_{\;\sigma\nu}-\Gamma^{\rho}_{\;\lambda\nu}\Gamma^{\lambda}_{\;\sigma\mu}=0\,, \\
T^{\rho}_{\;\mu\nu}&=\Gamma^{\rho}_{\;\nu\mu}-\Gamma^{\rho}_{\;\mu\nu}=0\,.
\end{align}
\end{subequations}
The non-metricity tensor is defined as
\begin{equation}
Q_{\alpha\beta\gamma}=\nabla_{\alpha}g_{\beta\gamma}\,,
\end{equation}
and we realize that it is symmetric in its two last indices. Thus, it admits two independent traces
\begin{equation}
Q_{\alpha}=Q_{\alpha\nu}^{\;\;\;\nu},\quad \bar{Q}_{\alpha}=Q_{\;\nu\alpha}^{\nu}\,.
\end{equation}
The affine connection in symmetric teleparallel gravity is related with the Levi-Civita connection through the decomposition
\begin{equation}\label{eq:conndec}
\Gamma^{\mu}_{\;\nu\rho}=\mathring{\Gamma}^{\mu}_{\;\nu\rho}+L^{\mu}_{\;\nu\rho}\,,
\end{equation}
where $L^{\mu}_{\;\nu\rho}$ is the disformation tensor
\begin{equation}
L^{\mu}_{\;\nu\rho}=\frac{1}{2}g^{\mu\lambda}(Q_{\lambda\nu\rho}-Q_{\nu\lambda\rho}-Q_{\rho\lambda\nu})\,.
\end{equation}
We also introduce the non-metricity scalar
\begin{equation}
Q=-\frac{1}{4}Q_{\alpha\mu\nu}Q^{\alpha\mu\nu}+\frac{1}{2}Q_{\alpha\mu\nu}Q^{\alpha\nu\mu}+\frac{1}{4}Q_{\alpha}Q^{\alpha}+\frac{1}{2}Q_{\alpha}\tilde{Q}^{\alpha}\,,
\end{equation}
and it follows from the decomposition~\eqref{eq:conndec} that it is related to the Ricci-scalar \(\mathring{R}\) of the Levi-Civita connection by
\begin{equation}\label{eq:riccidec}
\mathring{R} = -Q + B\,, \quad
B = \mathring{\nabla}_{\lambda}(Q^{\lambda}-\bar{Q}^{\lambda})\,.
\end{equation}
Because of the vanishing curvature and torsion, we can always find a local coordinate system in which the connection vanishes identically $\tilde{\Gamma}^{\mu}_{\;\nu\rho}=0$. This system is called  the coincident gauge \cite{BeltranJimenez:2017tkd,Hohmann:2021ast,Bahamonde:2022zgj}. As we will see, by employing this gauge, it is much easier to find the components of the connection in any other system of coordinates. The coordinate transformation of the connection from one coordinate system $\tilde{x}^{\mu}$ to an other $x^{\mu}$ is given by
\begin{equation}
\tilde{\Gamma}^{\mu}_{\;\nu\rho}=
\frac{\partial \tilde{x}^{\mu}}{ \partial x^{\mu}}\frac{\partial x^{\rho}}{ \partial \tilde{x}^{\rho}}\frac{\partial x^{\nu}}{ \partial \tilde{x}^{\nu}}\Gamma^{\mu}_{\;\nu\rho}-
\frac{\partial x^{\rho}}{ \partial \tilde{x}^{\rho}}\frac{\partial x^{\nu}}{ \partial \tilde{x}^{\nu}}\frac{\partial^{2}\tilde{x}^{\mu}}{\partial x^{\nu} \partial x^{\rho}}\,.
\end{equation}
If $\tilde{x}$ corresponds to the coincident gauge and $x^{\mu} = (t, r, \vartheta, \varphi)$, then the components of the connection in spherical coordinates are given by
\begin{equation}
\Gamma^{\mu}_{\;\nu\rho}=\frac{\partial x^{\mu}}{\partial \tilde{x}^{\sigma}}\frac{\partial^{2}\tilde{x}^{\sigma}}{\partial x^{\nu} \partial x^{\rho}}\,.
\end{equation}
We use the following ansatz
\begin{equation}
\tilde{x}^{0}=\tilde{t}(t,r),\quad \tilde{x}^{1}=\tilde{r}(t,r)\sin\vartheta\cos\varphi, \quad\tilde{x}^{2}=\tilde{r}(t,r)\sin\vartheta\sin\varphi,\quad \tilde{x}^{3}=\tilde{r}(t,r)\cos\vartheta\,,
\end{equation}
where $\tilde{t},\tilde{r}$ are some arbitrary functions of $t$ and $r$ that we will determine later in this section. We find the following non-zero components of the connection
\begin{gather}
\Gamma^{t}_{\;tt}=\frac{\partial_{r}\tilde{t}\;\partial_{tt}\tilde{r}-\partial_{r}\tilde{r}\;\partial_{tt}\tilde{t}}{\partial_{r}\tilde{t}\;\partial_{t}\tilde{r}-\partial_{r}\tilde{r}\;\partial_{t}\tilde{t}} , \quad
\Gamma^{t}_{\;tr}=\frac{\partial_{r}\tilde{t}\;\partial_{tr}\tilde{r}-\partial_{r}\tilde{r}\;\partial_{rt}\tilde{t}}{\partial_{r}\tilde{t}\;\partial_{t}\tilde{r}-\partial_{r}\tilde{r}\;\partial_{t}\tilde{r}}, \quad
\Gamma^{t}_{\;rr}=\frac{\partial_{r}\tilde{t}\;\partial_{rr}\tilde{r}-\partial_{r}\tilde{r}\;\partial_{rr}\tilde{t}}{\partial_{r}\tilde{t}\;\partial_{t}\tilde{r}-\partial_{r}\tilde{r}\;\partial_{t}\tilde{t}}\,,\nonumber \\
\Gamma^{r}_{\;tt}=\frac{\partial_{t}\tilde{t}\;\partial_{tt}\tilde{r}-\partial_{t}\tilde{r}\;\partial_{tt}\tilde{t}}{\partial_{r}\tilde{r}\;\partial_{t}\tilde{t}-\partial_{r}\tilde{t}\;\partial_{t}\tilde{r}}, \quad
\Gamma^{r}_{\;tr}=\frac{\partial_{t}\tilde{t}\;\partial_{tr}\tilde{r}-\partial_{t}\tilde{r}\;\partial_{rt}\tilde{t}}{\partial_{r}\tilde{r}\;\partial_{t}\tilde{t}-\partial_{r}\tilde{t}\;\partial_{t}\tilde{r}}, \quad
\Gamma^{r}_{\;rr}=\frac{\partial_{rr}\tilde{t}\;\partial_{t}\tilde{r}-\partial_{rr}\tilde{r}\;\partial_{t}\tilde{t}}{\partial_{r}\tilde{t}\;\partial_{t}\tilde{r}-\partial_{r}\tilde{r}\;\partial_{t}\tilde{r}}\,, \nonumber \\
\Gamma^{t}_{\;\vartheta\vartheta}=\frac{\tilde{r}\partial_{r}\tilde{t}}{\partial_{r}\tilde{r}\;\partial_{t}\tilde{t}-\partial_{r}\tilde{t}\;\partial_{t}\tilde{r}}, \quad
\Gamma^{r}_{\;\vartheta\vartheta}=\frac{\tilde{r}\partial_{t}\tilde{t}}{\partial_{r}\tilde{t}\;\partial_{t}\tilde{r}-\partial_{r}\tilde{r}\;\partial_{t}\tilde{t}}, \quad \Gamma^{t}_{\;\varphi\varphi}=\Gamma^{t}_{\;\vartheta\vartheta}\sin^{2}\vartheta, \quad
\Gamma^{r}_{\;\varphi\varphi}=\Gamma^{r}_{\;\vartheta\vartheta}\sin^{2}\vartheta\,, \nonumber \\
\Gamma^{\vartheta}_{\;t\vartheta}=\Gamma^{\varphi}_{\;t\varphi}=\frac{\partial_{t}\tilde{r}}{\tilde{r}}, \quad
\Gamma^{\vartheta}_{\;r\vartheta}=\Gamma^{\varphi}_{\;r\varphi}=\frac{\partial_{r}\tilde{r}}{\tilde{r}}, \quad
\Gamma^{\vartheta}_{\;\varphi\varphi}=-\cos\vartheta\sin\vartheta,\quad
\Gamma^{\varphi}_{\;\vartheta\varphi}=\cot\vartheta\,.
\end{gather}
A complete classification of all the metric affine spherically symmetric geometries has been done in \cite{Hohmann:2019fvf}.

We will now focus on the stationary case where the components of the connection depend only on $r$ and we will determine the general form of the functions $\tilde{r}(t,r)$ and $\tilde{t}(t,r)$ by direct integration of the components of the connection. We notice that it is quite straightforward to find the function $\tilde{r}(t, r)$ by integrating the component
\begin{equation}
\Gamma^{\vartheta}_{\;t\vartheta}=\frac{\partial_{t}\tilde{r}}{\tilde{r}}
\end{equation}
with respect to $t$. Thus, we obtain
\begin{equation}
\tilde{r}(t,r)=g_{1}(r)\;e^{\Gamma^{\vartheta}_{\;t\vartheta}t}\,,
\end{equation}
where $g_{1}(r)$ is an integration function. If we substitute this expression in the component $\Gamma^{\vartheta}_{\;r\vartheta}$, we find
\begin{equation}
\Gamma^{\vartheta}_{\;r\vartheta}=\frac{g_{1}'(r)}{g_{1}(r)}+(\Gamma^{\vartheta}_{\;t\vartheta})'t \,.
\end{equation}
Since there should be no time dependence, we should impose $(\Gamma^{\vartheta}_{\;t\vartheta})'=0$. Thus, $\Gamma^{\vartheta}_{\;t\vartheta}$ is constant. We set
\begin{equation}
\Gamma^{\vartheta}_{\;t\vartheta}=l\,,
\end{equation}
where $l$ is a real number. Then, we are left with
\begin{equation}
\Gamma^{\vartheta}_{\;r\vartheta}=\frac{g_{1}'(r)}{g_{1}(r)}\,.
\end{equation}
By integrating with respect to $r$, we find
\begin{equation}
g_{1}(r)=e^{\int\Gamma^{\vartheta}_{\;r\vartheta}dr+C_{a}}\,,
\end{equation}
where $C_{a}$ is an integration constant. We check that actually it does not affect the general form of the connection and thus, it can be set equal to zero without loss of generality. Then the general form of the function $\tilde{r}(t,r)$ can be written as
\begin{equation}
\tilde{r}(t,r)=e^{\int{f_{1}(r)}\;dr}\;e^{lt}\,,
\end{equation}
where $f_{1}(r)=\Gamma^{\vartheta}_{\;r\vartheta}$.

It takes some more work to determine the function $\tilde{t}(t, r)$. If we combine the components $\Gamma^{\vartheta}_{\;t\vartheta}$ and $\Gamma^{\vartheta}_{\;r\vartheta}$, we obtain
\begin{equation}
\label{int1}
\partial_{r}\tilde{t}=-\frac{\Gamma^{t}_{\;\vartheta\vartheta}}{\Gamma^{r}_{\;\vartheta\vartheta}}\partial_{t}\tilde{t}.
\end{equation}
If we combine the components
$\Gamma^{r}_{\;tt}$ and $\Gamma^{r}_{\;tr}$ and substitute (\ref{int1}), we get
\begin{equation}
\partial_{t}\tilde{t}=h(r)\tilde{t}\,,
\end{equation}
where
\begin{equation}
h(r)=f_{1}(r)\frac{l-\frac{\Gamma^{r}_{\;tt}}{\Gamma^{r}_{\;tr}}}{1+\frac{\Gamma^{t}_{\;\vartheta\vartheta}\Gamma^{r}_{\;tt}}{\Gamma^{r}_{\;\vartheta\vartheta}\Gamma^{r}_{tr}}}\,.
\end{equation}
By integrating with respect to $t$, we find an expression of the form
\begin{equation}
\tilde{t}=e^{h(r)t+g_{2}(r)}\,,
\end{equation}
where $g_{2}(r)$ is an arbitrary integration function of $r$. If we substitute this expression to the component $\Gamma^{r}_{\;tr}$, we obtain
\begin{equation}
l\;\Gamma^{r}_{\;\vartheta\vartheta}\left(-f_{1}(r)+th'(r)+\frac{h'(r)}{h(r)}+g'_{2}(r)\right)=\Gamma^{r}_{\;tr}\,.
\end{equation}
Since there should be no time dependence, we should set $h'(r)=0$. This implies that $h(r)=h=\textnormal{constant}$. We are then left with
\begin{equation}
g'_{2}(r)=\frac{\Gamma^{r}_{\;tr}}{l\;\Gamma^{r}_{\;\vartheta\vartheta}}+f_{1}(r)\,.
\end{equation}
One integration gives
\begin{equation}
g_{2}(r)=\int \left(\frac{\Gamma^{r}_{\;tr}}{l\;\Gamma^{r}_{\;\vartheta\vartheta}}+f_{1}(r)\right)dr+C_{b}\,,
\end{equation}
where $C_{b}$ is an integration constant. We check that it does not affect the general form of the connection. Thus, we set $C_{b}=0$ without loss of generality. The general form of the function $\tilde{t}(t,r)$ is given by
\begin{equation}
\tilde{t}(t,r)=e^{\int f_{2}(r)dr}e^{mt}\,,
\end{equation}
where $m=h$ and $ f_{2}(r)=\left(\frac{\Gamma^{r}_{\;tr}}{l\;\Gamma^{r}_{\;\vartheta\vartheta}}+f_{1}(r)\right)$\,.

Therefore, the components of the connection in a spherically symmetric and stationary spacetime take the form
\begin{gather}
\Gamma^{t}_{\;tt}=\frac{m^{2}f_{1}-l^{2}f_{2}}{mf_{1}-lf_{2}}, \quad \Gamma^{t}_{\;tr}=\frac{(l-m)f_{1}f_{2}}{lf_{2}-mf_{1}}, \quad
\Gamma^{t}_{\;rr}=\frac{f_{1}^{2}f_{2}+f_{2}f_{1}'-f_{1}(f_{2}^{2}+f_{2}')}{lf_{2}-mf_{1}},\nonumber \\
\Gamma^{r}_{\;tt}=\frac{lm(m-l)}{lf_{2}-mf_{1}}, \quad
\Gamma^{r}_{\;tr}=\frac{lm(f_{2}-f_{1})}{lf_{2}-mf_{1}}, \quad
\Gamma^{r}_{\;rr}=\frac{l(f_{2}^{2}+f_{2}')-m(f_{1}^{2}+f_{1}')}{lf_{2}-mf_{1}},\nonumber \\
\Gamma^{t}_{\;\vartheta\vartheta}=\frac{f_{2}}{mf_{1}-lf_{2}},\quad
\Gamma^{r}_{\;\vartheta\vartheta}=\frac{m}{lf_2-mf_1},\quad
\Gamma^{t}_{\;\varphi\varphi}=\Gamma^{t}_{\;\vartheta\vartheta}\sin^{2}\vartheta, \quad
\Gamma^{r}_{\;\varphi\varphi}=\Gamma^{r}_{\;\vartheta\vartheta}\sin^{2}\vartheta,
\nonumber \\
\Gamma^{\vartheta}_{\;t\vartheta}=\Gamma^{\varphi}_{\;t\varphi}, \quad
\Gamma^{\vartheta}_{\;r\vartheta}=\Gamma^{\varphi}_{\;r\varphi}, \quad
\Gamma^{\vartheta}_{\;\varphi\varphi}=-\cos\vartheta\sin\vartheta, \quad
\Gamma^{\varphi}_{\;\vartheta\varphi}=\cot\vartheta \,.
\end{gather}
Note that in order for the coincident gauge and thus also the connection to be well-defined, we must demand that the coordinate transformation is non-degenerate, and hence
\begin{equation}
\label{condition}
0 \neq \frac{1}{\tilde{t}\tilde{r}}\det\begin{pmatrix}
\partial_t\tilde{t} & \partial_r\tilde{t}\\
\partial_t\tilde{r} & \partial_r\tilde{r}
\end{pmatrix} = mf_1 - lf_2\,.
\end{equation}
By comparing with the paper \cite{DAmbrosio:2021zpm}, we realize that these components coincide with the second set with $c\rightarrow l$ and $k\rightarrow m-l$, while the first set arises as a subcase of the general connection for $m=0$. It is worthwhile to emphasize the importance of our result: We obtained the components of the connection by direct integration from the coincident gauge and our parametrization covers both sets without taking any limits. This is different from the approach discussed in~\cite{Bahamonde:2022zgj}, where the coordinate transformation to the coincident gauge is obtained by integrating the connection coefficients of the known spherically symmetric teleparallel connection. As a consequence, different parametrizations of the coincident gauge transformation are obtained for the two sets of connections, both of which are encompassed by our unified parametrization, which can be seen as follows. Considering first the case \(m = 0\), we have
\begin{equation}
\Gamma^r{}_{rr} = f_2 + \frac{f_2'}{f_2} = \left(\int f_2\,dr + \ln f_2\right)' = \left[\ln\left(f_2e^{\int f_2\,dr}\right)\right]' = \left\{\ln\left[\left(e^{\int f_2\,dr}\right)'\right]\right\}'\,,
\end{equation}
and thus
\begin{equation}
\tilde{t}(t,r) = e^{\int f_2\,dr} = \int e^{\int\Gamma^r{}_{rr}\,dr}dr\,,
\end{equation}
while \(\Gamma^{\varphi}{}_{r\varphi} = f_1\) and thus
\begin{equation}
\tilde{r}(t,r) = e^{lt + \int f_1\,dr} = e^{lt + \int f_1\,dr}\,.
\end{equation}
In the general case we have
\begin{equation}
\frac{\Gamma^t{}_{\vartheta\vartheta}}{\Gamma^r{}_{\vartheta\vartheta}} = -\frac{f_2}{m}\,,
\end{equation}
so that for \(m \neq 0\) we find\footnote{Note that there is a typo in~\cite[eq. (42a)]{Bahamonde:2022zgj}.}
\begin{equation}
\tilde{t}(t,r) = e^{mt + \int f_2\,dr} = \exp\left(mt - m\frac{\Gamma^t{}_{\vartheta\vartheta}}{\Gamma^r{}_{\vartheta\vartheta}}\,dr\right)\,,
\end{equation}
while
\begin{equation}
\frac{l\Gamma^t{}_{\vartheta\vartheta} + 1}{\Gamma^r{}_{\vartheta\vartheta}} = -f_1\,,
\end{equation}
leading to
\begin{equation}
\tilde{r}(t,r) = e^{lt + \int f_1\,dr} = \exp\left(lt - \int\frac{l\Gamma^t{}_{\vartheta\vartheta} + 1}{\Gamma^r{}_{\vartheta\vartheta}}dr\right)\,.
\end{equation}
In the following, we will continue with our unified parametrization in terms of \(l, m, f_{1,2}\), as it turns out to be more convenient for the task at hand.

\section{Newer General Relativity}
\label{sec:ngr}
The simplest symmetric teleparallel theory is described by the following gravitational action
\begin{equation}
S_{g}=-\frac{1}{2\kappa^{2}}\int \sqrt{-g}\;Q\;dx^{4}\,.
\end{equation}
We notice the similarity with the Einstein-Hilbert action where the Ricci scalar has been substituted by the non-metricity scalar using the relation~\eqref{eq:riccidec}, where the boundary term \(B\) has been omitted. It is worthy to emphasize that this theory is equivalent to GR and it is called the Symmetric Teleparallel Equivalent of GR (STEGR).

Newer GR is a quite simple generalization of STEGR. In this theory, the  non-metricity scalar is generalized to an arbitrary linear combination of its irreducible components
\begin{equation}
Q=c_{1}Q^{\mu\nu\rho}Q_{\mu\nu\rho}+c_{2}Q^{\mu\nu\rho}Q_{\rho\mu\nu}+c_{3}Q^{\rho\mu}_{\;\;\;\mu}Q_{\rho\nu}^{\;\;\;\nu}+c_{4}Q^{\mu}_{\;\;\mu\rho}Q_{\nu}^{\;\nu\rho}+c_{5}Q^{\mu}_{\;\mu\rho}Q^{\rho\nu}_{\;\;\;\nu} \,,
\end{equation}
where $c_{1},...,c_{5}$ are the parameters of the theory. If they have the following values
\begin{equation}
c_{1}=-\frac{1}{4}, \quad c_{2}=\frac{1}{2},\quad c_{3}=\frac{1}{4},\quad c_{4}=0, \quad c_{5}=-\frac{1}{2}\,,
\end{equation}
the theory is equivalent to STEGR and consequently to GR, up to a boundary term.

According to \cite{Flathmann:2020zyj}, there are two families of conditions for which the PPN parameters agree with those of GR. The following restrictions correspond to the first family
\begin{equation}
\label{parameters}
c_{1}=-\frac{1}{4}, \quad c_{2}=\frac{1}{2}\left(\frac{1}{2}+V\right),\quad c_{3}=\frac{1}{4},\quad c_{4}=\frac{1}{2}\left(\frac{1}{2}-V\right), \quad c_{5}=-\frac{1}{2}\,,
\end{equation}
where $V$ is a real number. Thus, we realize that if the three parameters have the STEGR values and the other two depend on each other, we are left with only one free parameter $V$. If the free parameter takes the value $V=1/2$, the theory reduces to STEGR. We could call this theory Type 1 Newer GR. Recently another Newer GR theory which is ghost-free was constructed based on a symmetry under transverse diffeomorphisms \cite{Bello-Morales:2024vqk}. In this theory, the following restrictions hold for the parameters
\begin{equation}
\label{parameterss}
c_{1}=-\frac{1}{4}, \quad c_{2}=\frac{1}{2},\quad c_{3}=\frac{1}{4}+V,\quad c_{4}=0, \quad c_{5}=-\frac{1}{2}\,,
\end{equation}
which belong to the second family from \cite{Flathmann:2020zyj}. $V$ is a real number. Similarly, four of the parameters have the STEGR values and the last one is free. If the free parameter takes the value $V=0$, the theory reduces to STEGR. We could call this theory Type 2 Newer GR. Thus, we realize that both theories have one free parameter.

Before we proceed with the derivation of the field equations, it is interesting to discuss how gravitational waves propagate in Newer GR since it appears that it does not depend on the value of the free parameter $V$. The propagation of gravitational waves in symmetric teleparallel theories has been already studied and all theories have been classified \cite{Hohmann:2018wxu}. There are four classes that correspond to different constraints on the parameters and consequently to different number and type of polarization modes. Considering the parameters in Newer GR \eqref{parameters} and \eqref{parameterss}, we realize that both sets satisfy the conditions
\begin{equation}
c_{2}+c_{4}+c_{5}=0 \;\textnormal{and}\; c_{5}\ne 0 \,.
\end{equation}
According to the aforementioned work, this implies that there are two polarization modes for any value of $V$ and these are the same modes that we find in GR as well.

The field equations can be derived by varying the action with respect to the metric only, since it turns out that variation with respect to the flat, symmetric connection yields a redundant equation~\cite{Hohmann:2021fpr,Hohmann:2022mlc}. Thus, we have \cite{Hohmann:2021ast}
\begin{eqnarray}
\kappa^{2}E_{\mu\nu}&=&-2\mathring{\nabla}_{\rho}\left[c_{1}Q^{\rho}_{\;\mu\nu}+c_{2}Q_{(\mu\nu)}^{\;\;\;\;\;\rho}+c_{3}Q^{\rho\sigma}_{\;\;\;\sigma}g_{\mu\nu}+c_{4}Q^{\sigma}_{\;\sigma(\mu}\delta^{\rho}_{\nu)}+\frac{1}{2}c_{5}(Q_{\sigma}^{\;\;\sigma\rho}g_{\mu\nu}+\delta^{\rho}_{(\mu}Q_{\nu)\sigma}^{\;\;\;\;\sigma})\right] \nonumber \\
&&\frac{1}{2}(c_{1}Q^{\rho\sigma\tau}Q_{\rho\sigma\tau}+c_{2}Q^{\rho\sigma\tau}Q_{\tau\rho\sigma}+c_{3}Q^{\tau\rho}_{\;\;\;\;\rho}Q_{\tau\sigma}^{\;\;\;\sigma}+c_{4}Q^{\rho}_{\;\;\rho\tau}Q_{\sigma}^{\;\sigma\tau}+c_{5}Q^{\rho}_{\;\rho\tau}Q^{\tau\sigma}_{\;\;\;\sigma})g_{\mu\nu}-c_{3}Q_{\mu\rho}^{\;\;\;\rho}Q_{\nu\sigma}^{\;\;\;\sigma}\nonumber \\
&&c_{1}(2Q^{\rho\sigma}_{\;\;\;\mu}Q_{\sigma\rho\nu}-Q_{\mu}^{\;\rho\sigma}Q_{\nu\rho\sigma}-2Q^{\rho\sigma}_{\;\;\;(\mu}Q_{\nu)\rho\sigma})+c_{2}(Q^{\rho\sigma}_{\;\;\;\mu}Q_{\sigma\rho\nu}-Q_{\mu}^{\;\rho\sigma}Q_{\nu\rho\sigma}-Q^{\rho\sigma}_{\;\;\;(\mu}Q_{\nu)\rho\sigma}) \nonumber \\
&&c_{4}\left[Q_{\rho}^{\;\rho\sigma}(Q_{\sigma\mu\nu}-2Q_{(\mu\nu)\sigma})+Q^{\rho}_{\;\rho\mu}Q^{\sigma}_{\;\sigma\nu}-Q^{\rho}_{\;\rho(\mu}Q_{\nu)\sigma}^{\;\;\;\;\sigma}\right]+\frac{1}{2}c_{5}\left[Q^{\rho\sigma}_{\;\;\;\sigma}(Q_{\rho\mu\nu}-2Q_{(\mu\nu)\rho})-Q_{\mu\rho}^{\;\;\;\rho}Q_{\nu\sigma}^{\;\;\;\sigma}\right]\,.
\end{eqnarray}
We continue our discussion by introducing the line element for a static spherically symmetric spacetime
\begin{equation}
ds^{2}=-A^{2}(r)\;dt^{2}+B^{2}(r)\;dr^{2}+r^{2}\;d\Omega^{2}\,,
\end{equation}
where \(d\Omega^2 = d\vartheta^2 + \sin^2\vartheta\,d\varphi^2\). In the spherically symmetric case, the non-zero components are $E_{tt},E_{rr},E_{rt}, E_{\vartheta\vartheta}$ and $E_{\varphi\varphi}=\sin^{2}\vartheta\;E_{\vartheta\vartheta}$. The expressions are quite lengthy and for this reason they can be found in the Appendix \ref{appendix1} and \ref{appendix2}. Due to the complexity of the equations, it is difficult to solve them in generally. By imposing some constraints on the free functions and parameters, we are able to find some interesting solutions. In the next section, we discuss these solutions.

\section{Solutions and their properties}
\label{sec:solutions}
In this section, we construct two solutions of the field equations in section~\ref{ssec:derivation}: One solution in Type 1 Newer GR and one solution which is common for both Type 1 and Type 2 Newer GR. We then discuss the possibility of them describing a wormhole in \ref{sub:worm}, their properties such as asymptotic flatness in \ref{sub:asym}, the Komar mass in \ref{sub:komar}, the gravitational redshift in \ref{sub:redshift}, the existence of singularities in \ref{sub:singularities}, the presence of horizons in \ref{sub:horizons}, the motion of test particles around these solutions in \ref{sub:geodesic}, the effect of light deflection in \ref{sub:light} and their causal structure in \ref{sub:structure}.

\subsection{Derivation of solutions}
\label{ssec:derivation}
In this subsection, we present the two solutions. The first one corresponds to the Type 1 Newer GR and the second one is common solution of both Type 1 and Type 2 theories.

\subsubsection{Solution 1 (in Type 1 Newer GR)}
We restrict ourselves to the special case
\begin{equation}
V=0\,,
\end{equation}
which provides a simple, non-trivial modification of STEGR. If we impose
\begin{equation}
f_{2}(r)=0 \;\textnormal{and} \;l=0 \,,
\end{equation}
we find one family of solutions with

\begin{equation}
\label{sol}
A(r) =1, \quad B(r)=\sqrt{rf_{1}(r)(2-rf_{1}(r))}\,,
\end{equation}
where we obtained $A(r)=1$ after a time rescaling of $A(r)=\;$constant. The choice of the function $f_{1}(r)$ will determine the properties of this solution. For reasons that will become apparent in the following subsections, we will study as a particular example the following function
\begin{equation}
\label{functionf1}
f_{1}(r)=\frac{1}{r-k}\,,
\end{equation}
where $k$ is a constant. For $k=0$, the solution reduces to Minkowski spacetime. $B(r)$ takes the following form
\begin{equation}
B(r)=\sqrt{\frac{r\;(r-2k)}{(r-k)^{2}}}\,.
\end{equation}
This implies that the radial coordinate $r\geq 0$ must further be restricted in order to satisfy $r\geq 2k\;$ and $r\neq k$, except in the special case \(k = 0\) corresponding to Minkowksi spacetime. We realize that this poses a restriction only for \(k > 0\), for which we must impose \(r \geq k\), while for \(k \leq 0\) the solution is valid in the whole coordinate range \(r \geq 0\).

\subsubsection{Solution 2 (in Type 1 and Type 2 Newer GR)}
In this section, we present a second solution that has a common form of the metric for both types of theories. By setting
\begin{equation}
m=0,\quad f_{1}(r)=0, \quad f_{2}(r)=k/r\,,
\end{equation}
in the Type 1 theory and
\begin{equation}
l=0,\quad f_{1}(r)=1/r, \quad f_{2}(r)=1/r\,,
\end{equation}
in the Type 2 theory, we find that the metric functions for this solution are given by
\begin{equation}
\label{sol2}
A(r)=qr \quad \text{and}\quad B(r)=B=\text{constant}\,,
\end{equation}
where $k, q$ are constants. In the first theory, the constants are related by the following conditions
\begin{equation}
\label{con1}
k =- \frac{2 q^{2}+4l^{2}}{q^{2}-l^{2}}, \quad V = 1/2 -\frac{q^{2} l^{2}}{2 q^{4} + q^{2} l^{2} - l^{4}}, \quad B=\pm\sqrt{\frac{4 q^{2} - 2l^{2}}{
 q^{2} - l^{2}}}
\end{equation}
This implies that we should have $q\ne\pm l$,  $q> l/\sqrt{2}$, $q< -l/\sqrt{2}$,   $q\ne 0$ and  $l\ne 0$. In the second theory,  the constants are related by the following conditions
\begin{equation}
\label{con2}
V = -\frac{q^{2}}{8m^{2}}, \quad B=\pm \sqrt{2}
\end{equation}
This implies that $m\ne 0$. It is worthy to mention that the constraints that arise do not allow us to consider the STEGR limit. The STEGR limit in Type 1 theory is given by $V=1/2$ and this implies that we should set $q=0$ or $l=0$ in (\ref{con1}). However,  $q=0$  would give $A=0$ while for $l=0$ the condition (\ref{condition}) would not be satisfied. The STEGR limit in Type 2 theory is given by $V=0$ and this implies that we should set $q=0$ in (\ref{con2}) which would give $A=0$. However, \(A = 0\) would imply that the metric is divergent, and so this solution does not have a counterpart in GR.

In the following subsections, we study the properties of these exotic objects and we mention when we make use of the function (\ref{functionf1}) in the case of the first solution.

\subsection{Possibility of traversable wormhole}
\label{sub:worm}
As the first property of the solutions given above, we will investigate whether they could represent a wormhole \cite{Morris:1988cz}. Let us assume that $r>0$ and $r_{0}$ is the wormhole throat with $r_{0}=r_{\text{min}}$. In the case of a wormhole, we can express the metric function $B(r)$ as
\begin{equation}
B(r)=\left(1-\frac{b(r)}{r}\right)^{-1/2}\,,
\end{equation}
where $b(r)$ is a function of r. With this parametrization, the so-called flaring out condition, which assures that the geometry is that of a wormhole, reads
\begin{equation}
b(r)<r_{0}<r, \quad b(r_{0})=r_{0}, \quad b'(r_{0})\leq 1 \,,
\end{equation}
where \(r_0\) denotes the radius of the throat.

\subsubsection{Solution 1} The term under the square root in (\ref{sol}) should be positive. This implies the conditions $f_{1}(r)>0$ and  $f_{1}(r)<2/r$ or $f_{1}(r)<0$ and  $f_{1}(r)>2/r$. However, for $r>0$ the second set of conditions cannot be satisfied, and so only the first set is valid. We solve (\ref{sol}) with respect to $f_{1}(r)$
\begin{equation}
f_{1}(r)=\frac{1\pm (1-B^{2})^{1/2}}{r}
\end{equation}
with $B^{2}\le 1$. For $B=1$, it is a Minkowksi solution. We also check that the conditions $f_{1}(r)>0$ and  $f_{1}(r)<2/r$ are satisfied.
The condition $B^{2}<1$ implies
\begin{equation}
b(r)<0\,.
\end{equation}
This condition is not satisfied on the throat of a wormhole, where we must have \(b(r_0) = r_0 > 0\). This means that this solution does not describe a wormhole. \\

\subsubsection{Solution 2}Since $B=$ constant, this solution cannot represent a wormhole because the flaring out condition is violated.

\subsection{Asymptotic flatness}
\label{sub:asym}
In this subsection, we investigate asymptotic flatness. The solution is asymptotic flat if we impose
\begin{equation}
\label{asflat}
\lim_{r\to\infty}A(r)=1,\; \lim_{r\to\infty} B(r) =1\,.
\end{equation}

\subsubsection{Solution 1} In this case, asymptotic flatness depends on the choice of the function  $f_1(r)$. The condition (\ref{asflat}) implies that $rf_{1}(r)\to 1$. There are several functions for which asymptotic flatness is satisfied and one of them is already introduced (\ref{functionf1}).

\subsubsection{Solution 2} It is easy to realize that this solution is not asymptotically flat.

\subsection{Komar mass}
\label{sub:komar}
The Komar mass of an object is an interesting quantity defined for stationary spacetimes and represents the mass of the object as computed by an observer at infinity \cite{Carroll:2004st, Poisson:2009pwt}. The Komar mass for any spherically symmetric spacetime is given by
\begin{equation}
\label{komar}
M=-\frac{1}{8\pi}\lim_{r\to\infty}\int_{0}^{\pi}\int_{0}^{2\pi}\frac{r^{2}g'_{tt}}{\sqrt{-g_{tt}g_{rr}}}\sin\vartheta\;d\varphi \;d\vartheta\,.
\end{equation}

\subsubsection{Solution 1}
In our case, we have $g_{tt}=\;$constant. Thus, the Komar mass vanishes. This implies that an observer perceives this object as massless.

\subsubsection{Solution 2}
We now calculate the Komar mass for the second solution. (\ref{komar}) takes the following form
\begin{eqnarray}
M&=&\frac{1}{8\pi}\lim_{r\to\infty}\int_{0}^{\pi}\int_{0}^{2\pi}\frac{2q^{2}r^{3}}{\sqrt{q^{2}r^{2}B^{2}}}\sin\vartheta\;d\varphi \;d\vartheta \notag \\
&=&\lim_{r\to\infty}\frac{qr^{2}}{B}=\infty.
\end{eqnarray}
The Komar mass of this exotic object becomes infinite, which is related to the fact that this solution is not asymptotically flat.

\subsection{Gravitational redshift}
\label{sub:redshift}
The gravitational redshift describes the change of the frequency of a light ray as measured by a stationary observer while it propagates to infinity and can be found by
\begin{equation}
\omega_{\infty}g_{tt\infty}=\omega g_{tt}\,,
\end{equation}
where $\omega$ and $\omega_{\infty}$ are the frequencies of a photon measured by a stationary observer at some radius $r$ and at infinity. $g_{tt}$ corresponds to the same radius $r$ as $\omega$.

\subsubsection{Solution 1}
This solution is asymptotically flat. For any value of \(r\), and thus also in the limit $r\to \infty$, we have $g_{tt}=-1$ and the gravitational redshift is given by \(\omega_{\infty}=\omega\). Thus, there is no gravitational redshift.

\subsubsection{Solution 2}
This solution is not asymptotically flat. Thus, for $r\to \infty$, we have $g_{tt}\to\infty$.The gravitational redshift vanishes, \(\omega_{\infty}=0\).

\subsection{Singularties}
\label{sub:singularities}
In order to check if there are any spacetime singularities, we compute the Kretschmann invariant
\begin{eqnarray}
\mathring{K}&=&\mathring{R}_{\alpha\beta\gamma\delta}\mathring{R}^{\alpha\beta\gamma\delta}\,.
\end{eqnarray}
\subsubsection{Solution 1} The Kretschmann invariant is given by
\begin{eqnarray}
\mathring{K}&=&\frac{4 \left(2 (r f_{1}-1)^{2} \left(rf'_{1}+f_{1}\right)^{2}+f_{1}^{2} (r f_{1}-2)^{2} (r f_{1}-1)^{4}\right)}{r^{6} \;f_{1}^{4}\; (r f_{1}-2)^{4}}\,.
\end{eqnarray}
The conditions for having spacetime singularities are the following
\begin{equation}
r=0, \quad f_{1}=0, \quad f_{1}=\frac{2}{r}\,.
\end{equation}
For the specific choice of function (\ref{functionf1}),  the Kretschmann invariant takes the following form
\begin{eqnarray}
\mathring{K}&=&\frac{4k^{4}(6k^{2}-8kr+3r^{2})}{r^{6}(r-2k)^{4}}
\end{eqnarray}
It is easy to check that there are two possible singularities at $r=0$ and $r=2k$ where the denominator of \(\mathring{K}\) vanishes. Their existence depends on the value of \(k\). For \(k < 0\), there is only one singularity at \(r = 0\), due to the condition \(r \geq 0\). Similarly, for \(k > 0\), there is only one singularity at \(r = 2k\) due to the condition \(r \geq 2k\). Finally, for the special case of Minkowski for which $k=0$, \(\mathring{K} = 0\) for any value of $r$ and thus, there are no singularities.

\subsubsection{Solution 2} The Kretschmann invariant is given by
\begin{eqnarray}
\mathring{K}&=&\frac{4 \left(B^4-2 B^2+3\right)}{B^4 r^4}\,.
\end{eqnarray}
Thus, there is one spacetime singularity at
\begin{equation}
r=0\,.
\end{equation}
We notice that \(B^4-2 B^2+3 \neq 0\) for any real value of \(B\), and so this singularity is present irrespective of the value of \(B\).

\subsection{Horizons}
\label{sub:horizons}
According to \cite{Poisson:2009pwt}, in static spacetimes the event horizon, the killing horizon and the apparent horizon coincide. The vector $t^{a} = \partial x^{a}/\partial t$ is a Killing vector in both spacetimes.

\subsubsection{Solution 1} Since $g_{\alpha\beta} t^{\alpha}t^{\beta}=$ constant, the vector is everywhere spacelike or timelike. Thus, there is no killing horizons. This implies that there are no event horizons or apparent horizons either.

\subsubsection{Solution 2} We have $g_{\alpha\beta} t^{\alpha}t^{\beta}=q^{2}r^{2}$ which is always positive. This implies that there is no Killing horizon and consequently no other horizons.

\subsection{Geodesic motion}
\label{sub:geodesic}
\label{subsec:orbits}

As we have already mentioned, we also study the motion of particles near these objects. In particular, we are interested in the radial motion and bound orbits. Due to spherical symmetry, the motion can be restricted at the equatorial plane $\vartheta=\pi/2$. The energy $E$ and the angular momentum $L$ are constants of the motion and they are given respectively by
\begin{eqnarray}
\label{EL}
E=A^{2}\dot{t}\,, \quad
L=r^{2}\dot{\varphi}\,,
\end{eqnarray}
where a dot indicates differentiation with respect to the proper time $\tau$ for massive particles or an affine parameter $\lambda$ for massless particles.  From the line element, we obtain
\begin{eqnarray}
\label{element}
&&s=-A^{2}\dot{t}^{2}+B^{2}\dot{r}^{2}+r^{2}\dot{\varphi}^{2}\,,
\end{eqnarray}
where $s=-1$ for the timelike case and $s=0$ for the null case. The last equation can be written in the form of a master equation
\begin{equation}
\label{rdot}
\frac{1}{2}\dot{r}^{2}+V_{\textnormal{eff}}(r)=0\,,
\end{equation}
where the effective potential $V_{\textnormal{eff}}(r)$ is given by
\begin{equation}
\label{potential}
V_{\text{eff}}(r)=-\frac{1}{2B^{2}}\left(s+\frac{E^{2}}{A^{2}}-\frac{L^{2}}{r^{2}}\right)\,.
\end{equation}
For inward motion, we have $\dot{r}<0$, while for outward motion, $\dot{r}>0$. If we use the expression for the energy (\ref{EL}) and substitute with $\dot{t}=\frac{dt}{dr}\frac{dr}{d\tau}$ (or $\dot{t}=\frac{dt}{dr}\frac{dr}{d\lambda}$), we find
\begin{equation}
\label{dtr}
\frac{dt}{dr}=\pm\;\frac{EB}{A^{2}}\left(s+\frac{E^{2}}{A^{2}}-\frac{L^{2}}{r^{2}}\right)^{-1/2}\,.
\end{equation}
In a similar manner and by using the expression for the angular momentum (\ref{EL}), we find
\begin{equation}
\label{dfr}
\frac{d\varphi}{dr}=\pm\frac{LB}{r^{2}}\left(s+\frac{E^{2}}{A^{2}}-\frac{L^{2}}{r^{2}}\right)^{-1/2}\,.
\end{equation}
The upper sign indicates outward motion and the bottom sign inward motion. The geodesics expressed in this way are convenient for computing effects such as the Shapiro time delay, the perihelion shift, the light deflection, the photon sphere and shadows.\\

\underline{Radial motion}:
In this paragraph, we restrict ourselves to particles that move along the radial direction. In that case, the angular momentum vanishes $L=0$. For the radial motion, it is interesting to investigate how a congruence of geodesics, that is a beam of particles evolves. In particular, we would like to study if the geodesics  converge or diverge and also how this convergence/divergence evolves with time. In order to do this, we have to compute the expansion scalar $\theta$ and derive the Raychaudhuri equation. The expansion of the congruence is given by
\begin{equation}
\theta=\mathring{\nabla}_{\alpha}u^{\alpha} \,,
\end{equation}
where the tangent vector field $u^{\alpha}$ to the geodesics is given by
\begin{equation}
u^{\alpha}=\frac{\partial x^{\alpha}}{\partial \tau}=(\dot{t}, \dot{r}, 0, 0)=\left(\frac{E}{A^{2}},\pm\frac{\sqrt{s+\frac{E^{2}}{A^{2}}}}{B},0,0\right)\,.
\end{equation}
The Raychaudhuri equations for a massive particle and a photon are given respectively \cite{Poisson:2009pwt}
\begin{eqnarray}
\frac{d\theta}{d\tau}&=&-\frac{1}{3}\theta^{2}-\sigma_{ab}\sigma^{ab}+\omega_{ab}\omega^{ab}-\mathring{R}_{\alpha\beta}u^{\alpha}u^{\beta}
\end{eqnarray}
\begin{eqnarray}
\frac{d\theta}{d\lambda}&=&-\frac{1}{2}\theta^{2}-\sigma_{ab}\sigma^{ab}+\omega_{ab}\omega^{ab}-\mathring{R}_{\alpha\beta}u^{\alpha}u^{\beta} \,,
\end{eqnarray}
where $\sigma_{\alpha\beta}$ is the shear tensor defined as
\begin{equation}
\sigma_{\alpha\beta}=\mathring{\nabla}_{(\alpha}u_{\beta)}-\frac{1}{n}\theta h_{\alpha\beta}
\end{equation}
and $\omega_{\alpha\beta}$ is the rotation tensor which is defined by
\begin{equation}
\omega_{\alpha\beta}=\mathring{\nabla}_{[\alpha}u_{\beta]}\,.
\end{equation}
$n$ takes the values $2$ or $3$ for null or timelike geodesics respectively  and $h_{\alpha\beta}$ is the transverse metric for which $h^{\alpha}_{\;\alpha}=n$ and $h_{\alpha\beta}u^{\alpha}=0$. At this point, we should clarify that the transverse metric is constructed in a different way for timelike and null worldlines. In the timelike case, it is expressed by
\begin{equation}
h_{\alpha\beta}=g_{\alpha\beta}+u_{\alpha}u_{\beta}\,.
\end{equation}
It is easy to find that it is given by
\begin{equation}
h_{\alpha\beta}=
\begin{pmatrix}
E^{2}-A^{2} &\mp EB\sqrt{\frac{E^{2}}{A^{2}}-1}& 0 & 0\\
\mp EB\sqrt{\frac{E^{2}}{A^{2}}-1} & \frac{B^{2}E^{2}}{A^{2}} & 0 & 0\\
0 & 0 & r^{2} & 0\\
0 & 0 & 0 & r^{2}\sin^{2}\theta
\end{pmatrix}
\end{equation}
In the null case, the transverse metric takes the following form
\begin{equation}
h_{\alpha\beta}=g_{\alpha\beta}+u_{\alpha}N_{\beta}+u_{\beta}N_{\alpha}\,,
\end{equation}
where we introduced a second null vector $N_{\alpha}$ for which $N_{\alpha}N^{\alpha}=0$, $N_{\alpha}u^{\alpha}=-1$ and $h_{\alpha\beta}N^{\alpha}=0$. This null vector is not unique and consequently, the null transverse metric is not unique. However, quantities that are constructed with this metric such as the expansion are unique. It is easy to check that the covector $N_{\alpha}=\{-A^{2}/2E, \mp AB/2E,0,0\}$ satisfies the aforementioned conditions. Thus, the null transverse metric constructed with this vector is given by
\begin{equation}
h_{\alpha\beta}=\mathrm{diag}\{0,0, r^{2}, r^{2}\sin^{2}\vartheta\}\,,
\end{equation}
where we have assumed that for each case $A>0$.
  \\~\\
\underline{Bound orbits}:
It is also interesting to study the motion of a particle when it follows a bound orbit around the object. We should find  two radii \(r_{1,2}\) for which
\begin{equation}
\dot{r}=0, \quad V_{\text{eff}}(r) = 0
\end{equation}
and for $r_{1} < r < r_{2}$
\begin{equation}
V_{\text{eff}}(r) < 0\,.
\end{equation}
In this case the test particle oscillates between \(r_{1,2}\). A special case is when \(r_1 = r_2\), which implies \(V_{\text{eff}}'(r_{1,2}) = 0\), and thus \(\ddot{r} = 0\). In this case the orbit is circular. It is stable for \(V_{\text{eff}}''(r_{1,2}) > 0\), since then the potential has a minimum. Thus, the extra conditions for bound orbits to be circular are
\begin{equation}
\ddot{r}=0 \quad \textnormal{and} \quad \frac{d V_{\text{eff}}}{dr}=0\,.
\end{equation}

\begin{figure}
\includegraphics[width=0.5\textwidth]{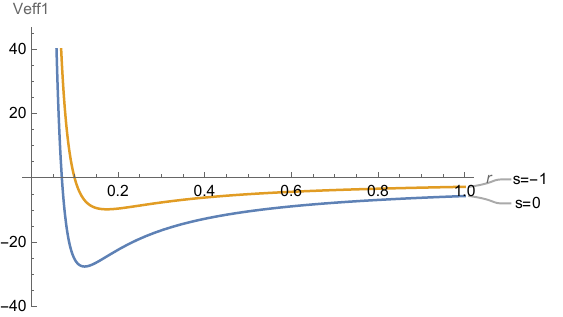}
\caption{The effective potential for the first solution for massive and massless particles with the fixed values $E=\sqrt{2}, L=0.1$ and $k=-10$. For massive particles $r_{c}=0.1$ while for massless particles $r_{c}=0.05$.}
\label{Veff1}
\end{figure}
\begin{figure}
\includegraphics[width=0.5\textwidth]{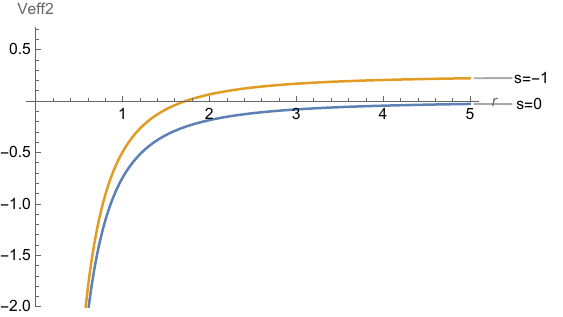}
\caption{The effective potential for the second solution for massive and massless particles with the fixed values $E=2, L=1, q=1$ and $B=\sqrt{2}$. For massive particles $r_{c}=\sqrt{3}$ while for massless particles there is no turning point.}
\label{Veff2}
\end{figure}

\subsubsection{Solution 1} We discuss the radial motion first. The expansion, the rotation and the quantity$R_{ab}u^{a}u^{b}$ take the following form
\begin{eqnarray}
&\theta=\pm\frac{2\sqrt{E^{2}+s}\;}{rB}\,,& \\ \notag\\
&\omega_{ab}=0\,,&\\ \notag\\
&R_{ab}u^{a}u^{b}=\frac{2 B' \left(E^2+s\right)}{B^3 r}\,.&
\end{eqnarray}
The shear tensor for the timelike and null case takes the following form respectively
\begin{equation}
\sigma_{\alpha\beta (t)}=
\begin{pmatrix}
\mp\frac{2 \left(E^2-1\right)^{3/2}}{3 B r} & \frac{2 E \left(E^2-1\right)}{3 r} & 0 & 0 \\
\frac{2 E \left(E^2-1\right)}{3 r} & \mp\frac{2 B E^2 \sqrt{E^2-1}}{3 r} &0 &0 \\
0 & 0& \pm\frac{r\sqrt{E^2-1} }{3 B} &0 \\
0 & 0 & 0 & \pm\frac{r\sqrt{E^2-1} }{3 B}\sin^{2}\vartheta
\end{pmatrix}\,,
\end{equation}
\begin{equation}
\sigma_{\alpha\beta (n)}=0\,.
\end{equation}
We realize that $\theta<0$ and $\theta>0$ for inward and outward motion respectively since $B>0$. We also notice that in the timelike case, we should have $E^{2}+s\ge0$. The fact that $\theta$ is not negative for all cases indicates that there are no trapped surfaces and this is another proof of the absence of apparent horizons. Positive expansion implies that the geodesic diverge while negative expansion implies that the geodesics converge. The Raychaudhuri equation for the timelike and null cases both for inward and outward motion takes the following form respectively
\begin{eqnarray}
\frac{d\theta}{d\tau}=-\frac{2 \left(E^2-1\right) \left(r B'+B\right)}{B^3 r^2}=-\frac{2 \left(E^2-1\right)(r-k)  \left(3 k^2-3 k r+r^2\right)}{r^3 (r-2 k)^{2}}\,,
\end{eqnarray}
\begin{eqnarray}
\frac{d\theta}{d\lambda}=-\frac{2E^{2} \left(r B'+B\right)}{B^3 r^2}=-\frac{2 E^2 (r-k)  \left(3 k^2-3 k r+r^2\right)}{r^3 (r-2 k)^{2}}\,.
\end{eqnarray}
We realize that for the specific function $f_{1}$ (\ref{functionf1}), $\frac{d\theta}{d\lambda}<0$ as well as $\frac{d\theta}{d\tau}\le0$ since we consider $E^{2}+s\ge0$. This implies that the initially converging beam will converge more rapidly as time passes while the initially diverging beam will converge less rapidly. Thus, the expansion or the volume of the beam decreases and the worldlines will cross at a caustic point. These results are consistent with General Relativity, the focusing theorem and consequently the attractive character of gravity. Only in the timelike case, the equality implies that there is no change in the evolution of the expansion.

In order to discuss bound orbits, we first express the effective potential for this solution
\begin{equation}
V_{\text{eff1}}(r)=-\frac{(r-k)^{2}}{2r(r-2k)}\left(s+E^{2}-\frac{L^{2}}{r^{2}}\right)\,.
\end{equation}
We notice that due to the constraints appearing in the case $k>0$, we must impose $r> 2k$ and it is not possible to have $r=k$.
We find one radius for which $V_{\text{eff1}}(r)=0$ which is given by
\begin{equation}
	\label{radius}
r_{c}=\frac{L}{\sqrt{E^{2}+s}}\,,
\end{equation}
with $E^{2}+s>0$. The existence of only one zero crossing implies that there are no bound orbits. Since $V_{\text{eff}1}(r) < 0$ for $r > r_{c}$,  if the particle is approaching the object, then it is scattered to infinity if $r_{c}>2k$ and the point $r_{c}$ corresponds to the nearest point of its orbit to the object. Otherwise for $r_{c}<2k$ it encounters the singularity at $r=2k$.

\subsubsection{Solution 2} We start with the radial motion. The expansion, the rotation and the quantity $R_{ab}u^{a}u^{b}$ take the following form
\begin{eqnarray}
&\theta=\pm\frac{2 E^2+3 q^2 r^2 s}{B q^2 r^3 \sqrt{\frac{E^2}{q^2 r^2}+s}}\,,& \\ \notag\\
&\omega_{ab}=0\,,&\\ \notag\\
&R_{ab}u^{a}u^{b}=\frac{2 E^2}{B^2 q^2 r^4}\,.&
\end{eqnarray}
The shear tensor for the timelike and null case takes the following form respectively
\begin{equation}
\sigma_{\alpha\beta(t)}=
\begin{pmatrix}
\mp\frac{2 E^2 \sqrt{\frac{E^2}{q^2 r^2}-1}}{3 B r}&\frac{2 E^3}{3 q^2 r^3} & 0 & 0 \\
\frac{2 E^3}{3 q^2 r^3} &\mp\frac{2 B E^4}{3 q^4 r^5 \sqrt{\frac{E^2}{q^2 r^2}-1}}&0 &0 \\
0 & 0& \pm\frac{E^2}{3 B q^2 r \sqrt{\frac{E^2}{q^2 r^2}-1}} &0 \\
0 & 0 & 0 & \pm\frac{E^2}{3 B q^2 r \sqrt{\frac{E^2}{q^2 r^2}-1}}\sin^{2}\vartheta\,,
\end{pmatrix}
\end{equation}
\begin{equation}
\sigma_{\alpha\beta(n)}=
\begin{pmatrix}
\mp\frac{Eq}{B} &\frac{E}{r} & 0 & 0 \\
\frac{E}{r} & \mp\frac{BE}{qr^{2}} &0 &0 \\
0 & 0&0 &0 \\
0 & 0 & 0 & 0
\end{pmatrix}\,.
\end{equation}
Assuming $B>0$, we notice that $\theta<0$ and $\theta>0$ for inward and outward motion respectively for both the timelike and null case because $E^{2}-q^{2}r^{2}>0$. The non-existence of trapped surfaces is also implied. The Raychaudhuri equations for massive and massless particles both for inward and outward motion are given respectively by
\begin{eqnarray}
\frac{d\theta}{d\tau}=-\frac{4E^4-6 E^2 q^2 r^2+3 q^4 r^4}{B^2 q^2 r^4 \left(E^2-q^2 r^2\right)}\,,
\end{eqnarray}
\begin{eqnarray}
\frac{d\theta}{d\lambda}&=&-\frac{4E^{2}}{ B^{2} q^{2}r^{4}}\,.
\end{eqnarray}
We realize that both $\frac{d\theta}{d\tau}$, $\frac{d\theta}{d\lambda}$ are negative. These results are consistent with the ones of General Relativity and the focusing theorem.

In order to discuss bound orbits, we provide the expression for the effective potential for this solution
\begin{equation}
V_{\text{eff2}}(r)=-\frac{1}{2B^{2}}\left(s+\frac{E^{2}}{q^{2}r^{2}}-\frac{L^{2}}{r^{2}}\right)\,.
\end{equation}
In this case, we do not find bound orbits for photons since there is no radius for which $V_{\text{eff2}}(r)=0$ while $s=0$. We find one radius for which $V_{\text{eff2}}(r)=0$ in the timelike case
\begin{equation}
r_{c}=\sqrt{\frac{L^{2}-\frac{E^{2}}{q^{2}}}{s}}\,,
\end{equation}
with $E^{2}/q^{2}-L^{2}>0$. We emphasize that this holds only for $s=-1$. The existence of only one zero crossing does not indicate bound orbits. Since $V_{\text{eff}2}(r)<0$ for $r<r_{c}$, the particle is confined to the region $r < r_{c}$ and if it is initially moving outward, it will turn around at $r_{c}$ and then continues falling towards the singularity at $r=0$.
\\

\subsection{Light deflection}
\label{sub:light}
In this subsection, we would like to investigate the shape of the orbit of light when it passes near the objects we discuss. In particular, we would like to compute the deflection angle, the angle about which the light is bent. This is given by
\begin{eqnarray}
\label{deflection}
\Delta \phi&=&\int_{r_{c}}^{r_{S}}\frac{d\phi}{dr}dr+\int_{r_{c}}^{r_{R}}\frac{d\phi}{dr}dr+\Psi_{R}-\Psi_{S}\nonumber \\
&=&\int_{r_{c}}^{r_{S}}\frac{LB}{r^{2}}\left(\frac{E^{2}}{A^{2}}-\frac{L^{2}}{r^{2}}\right)^{-1/2}dr+\int_{r_{c}}^{r_{R}}\frac{LB}{r^{2}}\left(\frac{E^{2}}{A^{2}}-\frac{L^{2}}{r^{2}}\right)^{-1/2}dr+\Psi_{R}-\Psi_{S} \,.
\end{eqnarray}
where $\Psi_{S}$ and $\Psi_{R}$ are the angles at which the light is emitted at $r=r_{S}$ and received at $r=r_{R}$ respectively. We can calculate $\Psi_{R}-\Psi_{S}$ according to \cite{Ishihara:2016vdc}. For asymptotically flat spacetimes, we have $\Psi_{R}-\Psi_{S}=-\pi$. The turning point $r_{c}$ can be found by solving the equation $\frac{dr}{d\lambda}=0$.

\subsubsection{Solution 1}
If we try to calculate the expression (\ref{deflection}), we realize that we obtain a lengthy expression that does not provide much insight. We will use an alternative method in order to derive a meaningful result for the light deflection at large distances. First, we find that the turning point is given by
\begin{equation}
\label{rc}
r_{c}=\frac{L}{E}\,.
\end{equation}
We use (\ref{rc}) so our expression will not involve $E$ and $L$. Then we replace the radii $r_{S}, r_{R}$ and $r_{c}$ with $r_{S}/\epsilon, r_{R}/\epsilon$ and $r_{c}/\epsilon$ respectively where $\epsilon$ is infinitesimally small.
\begin{eqnarray}
\Delta \phi=\int_{r_{c}/\epsilon}^{r_{S}/\epsilon}\frac{r_{c}B}{\epsilon r^{2}}\left(1-\frac{r_{c}^{2}}{\epsilon^{2}r^{2}}\right)^{-1/2}dr+\int_{r_{c}/\epsilon}^{r_{R}/\epsilon}\frac{r_{c}B}{\epsilon r^{2}}\left(1-\frac{r_{c}^{2}}{\epsilon^{2}r^{2}}\right)^{-1/2}dr+\Psi_{R}-\Psi_{S} \,.
\end{eqnarray}
We now take the limit for $r_{S}, r_{R}\to \infty$
\begin{eqnarray}
\label{dflimit}
\Delta\phi_{r_{S}, r_{R}\to \infty}&=&\frac{8 r_{c}^{5/2} K\left(\frac{4 \epsilon k}{r_{c}+2 \epsilon k}\right)}{(r_{c}-\epsilon k) (r_{c}+\epsilon k) \sqrt{r_{c}+2 \epsilon k}}-\frac{8 r_{c}^{3/2} F\left(\frac{\pi }{4}|\frac{4 \epsilon k}{r_{c}+2 \epsilon k}\right)}{(r_{c}+\epsilon k) \sqrt{r_{c}+2 \epsilon k}}+\frac{4 r_{c}^{3/2} \epsilon k \Pi \left(\frac{2 \epsilon k}{r_{c}+\epsilon k}|\frac{4 \epsilon k}{r_{c}+2 \epsilon k}\right)}{(r_{c}-\epsilon k) (r_{c}+\epsilon k) \sqrt{r_{c}+2 \epsilon k}} \notag \\
&&-\frac{4 r_{c}^{5/2} \Pi \left(\frac{2 \epsilon k}{r_{c}+\epsilon k}|\frac{4 \epsilon k}{r_{c}+2 \epsilon k}\right)}{(r_{c}-\epsilon k) (r_{c}+\epsilon k) \sqrt{r_{c}+2 \epsilon k}}+\frac{4 r_{c}^{3/2} \Pi \left(\frac{2 \epsilon k}{r_{c}+\epsilon k};\frac{\pi }{4}|\frac{4 \epsilon k}{r_{c}+2 \epsilon k}\right)}{(r_{c}+\epsilon k) \sqrt{r_{c}+2 \epsilon k}}-\frac{8 \sqrt{r_{c}} \epsilon^2 k^2 K\left(\frac{4 \epsilon k}{r_{c}+2 \epsilon k}\right)}{(r_{c}-\epsilon k) (r_{c}+\epsilon k) \sqrt{r_{c}+2 \epsilon k}}\notag \\
&&-\frac{8 \sqrt{r_{c}} \epsilon k F\left(\frac{\pi }{4}|\frac{4 \epsilon k}{r_{c}+2 \epsilon k}\right)}{(r_{c}+\epsilon k) \sqrt{r_{c}+2 \epsilon k}}+\Psi_{R}-\Psi_{S}\,,
\end{eqnarray}
where $F, K$ and $\Pi$ are all elliptic integrals. Their expressions are given by
\begin{equation}
F(\phi\;|\;k^{2})=\int_{0}^{\phi}\frac{d\theta}{\sqrt{1-k^{2}\sin^{2}\theta}}, \quad K(k)=F\left(\frac{\pi}{2}\;|\;k^{2}\right), \quad
\Pi(n;\phi\;|\;m)=\int_{0}^{\sin\phi}\frac{1}{1-nt^{2}}\frac{dt}{\sqrt{(1-mt^{2})(1-t^{2})}}\,.
\end{equation}
We consider a Taylor expansion of (\ref{dflimit}) around $\epsilon$ and we check if some low order term vanishes. The last expression correspond to the zeroth order term of the expansion. It is easy to check that for $\epsilon \to 0$, the sum of the terms that include elliptic functions gives $\pi$. If we also consider the term $\Psi_{R}-\Psi_{S}=-\pi$, in total we get zero. We then take the first derivative of (\ref{dflimit}) with respect to $\epsilon$. We realize that it does not vanish for $\epsilon \to 0$  and actually the limit is of the form $0/0$. We apply the Hopital's rule for this limit and we conclude that it vanishes. We can consider then that higher order terms are negligible and thus, the light deflection in large distances vanishes which agress with the result for a flat spacetime.

\subsubsection{Solution 2}
For this solution, there is no turning point. This implies that there is no deflection and that the particle hits the singularity at $r=0$ even if initially $\dot{r}\ne0$. In order to prove this, we will express $t$ and $r$ with respect to the affine parameter $\lambda$ by integrating (\ref{EL}) and (\ref{rdot}). For the second solution we have $A=qr$ and for massless particles $s=0$. Thus, (\ref{rdot}) reduces to
\begin{equation}
\frac{dr}{d\lambda}=\frac{1}{Br}\left(\frac{E^{2}}{q^{2}}-L^{2}\right)^{1/2}\,.
\end{equation}
We assume that for $r=r_{0}=0$, $\lambda=\lambda_{0}$ where $\lambda_{0}$ a finite value. By integrating, we obtain
\begin{equation}
r^{2}(\lambda)=\frac{2}{B}\left(\frac{E^{2}}{q^{2}}-L^{2}\right)^{1/2}(\lambda-\lambda_{0})\,.
\end{equation}
We use this equation to derive an expression for $t$ with respect to $\lambda$. (\ref{EL}) takes the following form
\begin{equation}
\frac{dt}{d\lambda}=\frac{E}{q^{2}r^{2}(\lambda)}\,.
\end{equation}
One integration gives
\begin{equation}
t(\lambda)=\frac{EB}{2q^{2}}\left(\frac{E^{2}}{q^{2}}-L^{2}\right)^{-1/2}\ln(\lambda-\lambda_{0})+t_{0}\,.
\end{equation}
We realize that the light ray hits the singularity $r=0$ for a finite value of the parameter $\lambda_{0}$ but at infinite time $t\to-\infty$. Thus, there is not deflection in that case.

\subsection{Causal structure}
\label{sub:structure}
In order to better comprehend the solutions we presented and their properties, we discuss their causal structure in this section by constructing the Penrose diagrams. We should first find a suitable coordinate transformation $r=r(\rho)$ that brings the metric into the following form
\begin{equation}
ds^{2}=-F^{2}dt^{2}+F^{-2}d\rho^{2}+r^{2}(\rho)d\Omega^{2}\,,
\end{equation}
where $F=F(r(\rho))$. Thus, we have $A=F$ and we should solve the equation
\begin{equation}
\label{eqAB}
B\;dr=\frac{1}{F}\;d\rho\,.
\end{equation}
Then, we find the tortoise coordinate $\rho^{*}$ by
\begin{equation}
\label{tc}
\rho^{*}=\int_{0}^{\rho} \frac{d\rho'}{F^{2}}\,,
\end{equation}
and we introduce the null coordinates
\begin{equation}
\label{ct1}
u=t-\rho^{*},\quad U=t+\rho^{*}\,.
\end{equation}
In order to restrict the coordinates into a final range, we use the following coordinate transformation
\begin{equation}
\label{ct2}
\tilde{u}=\tanh u,\quad \tilde{U}=\tanh U\,.
\end{equation}
For the Penrose diagrams, we use the coordinates
\begin{equation}
\label{ct3}
T=\frac{\tilde{U}+\tilde{u}}{2},\quad L=\frac{\tilde{U}-\tilde{u}}{2}\,.
\end{equation}
The metric then takes the following form
\begin{equation}
ds^{2}=-F^{2}\frac{dL^{2}-dT^{2}}{(1-(T-L)^{2})(1-(T+L)^{2})}\,.
\end{equation}
We go into more details for each solution in the next paragraphs.

\subsubsection{Solution 1} The first solution is given by (\ref{sol}) and we use the specific function (\ref{functionf1}). Solving the equation (\ref{eqAB}), we obtain
\begin{equation}
\label{rho}
\rho=\sqrt{r(r-2k)}-2k\arctan\frac{k-r+\sqrt{r(r-2k)}}{k}\,.
\end{equation}
We compute the tortoise coordinate by (\ref{tc})
\begin{equation}
\rho^{*}=\frac{\rho}{A^{2}}\,.
\end{equation}
In order to draw the Penrose diagram, we express the coordinates $T,L$ with respect to $t$ and $r$
\begin{eqnarray}
T=\frac{1}{2}(\tanh(t+\rho(r)/A^{2})+\tanh(t-\rho(r)/A^{2}))\,, \notag \\
L=\frac{1}{2}(\tanh(t+\rho(r)/A^{2})-\tanh(t-\rho(r)/A^{2}))\,,
\end{eqnarray}
where $\rho(r)$ is given by $(\ref{rho})$.

As we have discussed, there is one singularity in this spacetime, whose location depends on the parameter \(k\). The first case is found for $k<0$ at $r\to 0$ and we have $\rho\to -k\pi/2$ which corresponds to a point
\begin{equation}
T=\frac{1}{2}(\tanh(t-k\pi/2A^{2})+\tanh(t+k\pi/2A^{2})),\quad L=\frac{1}{2}(\tanh(t-k\pi/2A^{2})-\tanh(t+k\pi/2A^{2}))
\end{equation}
on the Penrose diagram while the second is found for $k>0$ at $r\to 2k$ and we have $\rho\to k\pi/2$ which corresponds to the point
\begin{equation}
T=\frac{1}{2}(\tanh(t-k\pi/2A^{2})+\tanh(t+k\pi/2A^{2})),\quad L=-\frac{1}{2}(\tanh(t-k\pi/2A^{2})-\tanh(t+k\pi/2A^{2}))
\end{equation}
Light rays reach the singularities in finite coordinate time. For $r\to \infty$, we have $\rho\to \infty$ and thus, $T\to 0$, $L\to 1$. For $t\to 0$, we have $T\to 0$ and $L\to \tanh(\rho/A^{2})$ while for $t\to \infty$, we have $T\to 1$, $L\to 0$.

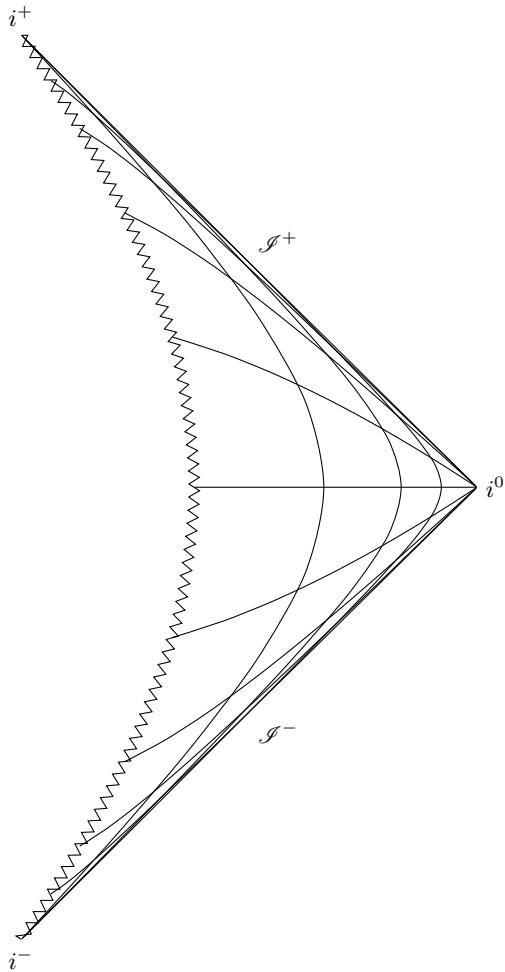
\begin{figure}
\tikzsetnextfilename{penrose1}
\begin{tikzpicture}[scale=3,/pgf/fpu/install only={reciprocal}]
\foreach \y in {-1.6,-1.2,-0.8,-0.4,0,0.4,0.8,1.2,1.6} \draw[domain=0.4:4,smooth,variable=\x,samples=25] plot ({tanh(\x + \y) + tanh(\x - \y)},{tanh(\x + \y) - tanh(\x - \y)});
\foreach \x in {0.8,1.2,1.6} \draw[domain=-4:4,smooth,variable=\y,samples=25] plot ({tanh(\x + \y) + tanh(\x - \y)},{tanh(\x + \y) - tanh(\x - \y)});
\draw[domain=-4:4,smooth,variable=\y,samples=25,decorate,decoration={zigzag,amplitude=2,segment length=5}] plot ({tanh(0.4 + \y) + tanh(0.4 - \y)},{tanh(0.4 + \y) - tanh(0.4 - \y)});
\draw (0,2) node [above] {$i^+$};
\draw (0,-2) node [below] {$i^-$};
\draw (2,0) node [right] {$i^0$};
\draw (1,1) node [above right] {$\mathscr{I}^+$};
\draw (1,-1) node [below right] {$\mathscr{I}^-$};
\end{tikzpicture}
\caption{Conformal diagram of the first solution. The left edge shows the singularity at \(r = 0\) in the case \(k < 0\), or \(r = 2k\) in the case \(k > 0\), which is reached in finite coordinate time.}
\label{fig:penrose1}
\end{figure}

\subsubsection{Solution 2}
The second solution is given by (\ref{sol2}). Solving the equation (\ref{eqAB}), we obtain
\begin{equation}
\rho=\frac{Bq}{2}r^{2}
\end{equation}
We also compute the tortoise coordinate by (\ref{tc})
\begin{equation}
\rho^{*}=\frac{B}{2q}ln\left(\frac{2}{Bq}\rho\right)\,.
\end{equation}
In order to draw the Penrose diagram, we express the coordinates $T,L$ with respect to $t$ and $r$
\begin{eqnarray}
T=\frac{1}{2}(\tanh(t+B\ln r/q)+\tanh(t-B\ln r/q))\,, \notag \\
L=\frac{1}{2}(\tanh(t+B\ln r/q)-\tanh(t-B\ln r/q))\,.
\end{eqnarray}

It is worthy to emphasize that the shape of the Penrose diagram reminds the Penrose diagram for a Mikowkski spacetime. However, they are not the same. For the Minkowski spacetime the vertical axis corresponds to $r=0$ while in our case, $r=0$ is a singularity that corresponds to the point $L=-1$ and $T=0$ for finite $t$. However, light rays do not reach it in finite coordinate time but only approach asymptotically as time goes to infinity. We can check that for $r\to \infty$, we have $T\to 0$ and $L\to  1$. For $r=$constant and $t\to 0$, we have $T\to 0$ and $L\to \tanh(B\ln r/q)$ while for $t\to \infty$, we have $T\to 1$, $L\to 0$.

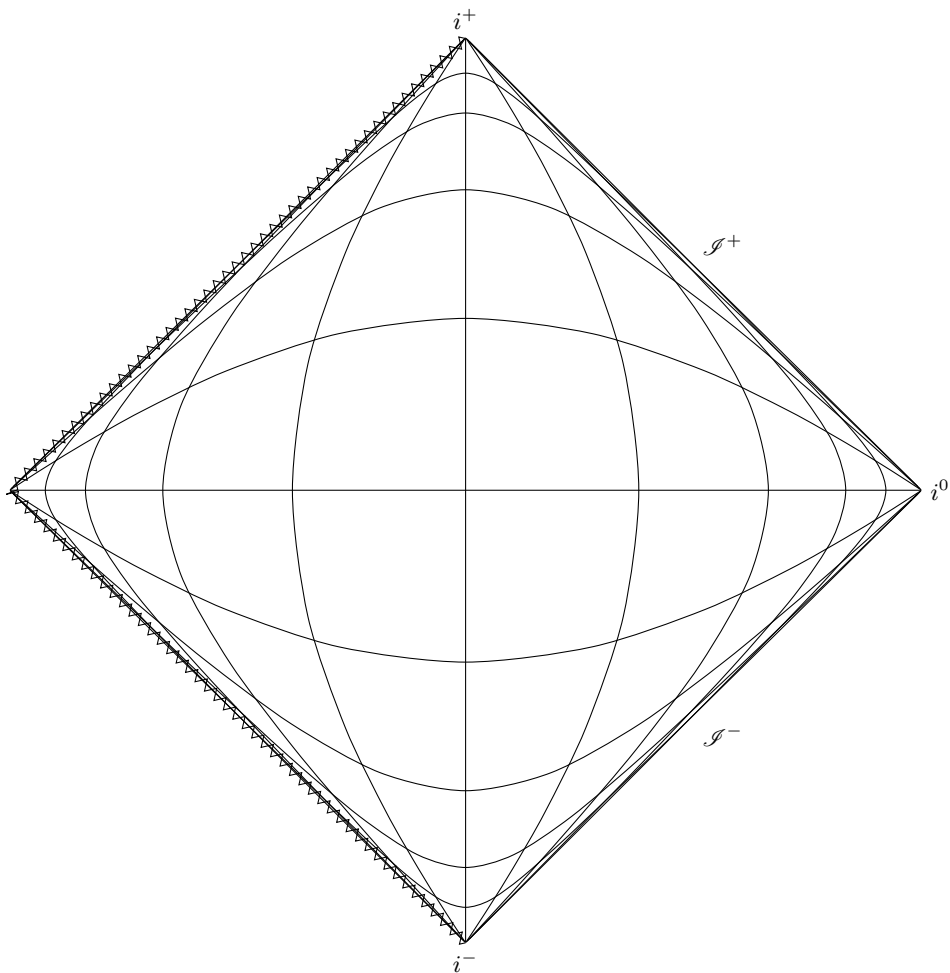
\begin{figure}
\tikzsetnextfilename{penrose2}
\begin{tikzpicture}[scale=3,/pgf/fpu/install only={reciprocal}]
\foreach \y in {-1.6,-1.2,-0.8,-0.4,0,0.4,0.8,1.2,1.6} \draw[domain=-4:4,smooth,variable=\x,samples=25] plot ({tanh(\x + \y) + tanh(\x - \y)},{tanh(\x + \y) - tanh(\x - \y)});
\foreach \x in {-1.6,-1.2,-0.8,-0.4,0,0.4,0.8,1.2,1.6} \draw[domain=-4:4,smooth,variable=\y,samples=25] plot ({tanh(\x + \y) + tanh(\x - \y)},{tanh(\x + \y) - tanh(\x - \y)});
\draw (0,2) node [above] {$i^+$};
\draw (0,-2) node [below] {$i^-$};
\draw (2,0) node [right] {$i^0$};
\draw (1,1) node [above right] {$\mathscr{I}^+$};
\draw (1,-1) node [below right] {$\mathscr{I}^-$};
\draw[decorate,decoration={zigzag,amplitude=2,segment length=5}] (0,-2) -- (-2,0) -- (0,2);
\end{tikzpicture}
\caption{Conformal diagram of the second solution. The left edges correspond to the singularity at \(r = 0\), and we see that it cannot be reached in finite coordinate time.}
\label{fig:penrose2}
\end{figure}

\section{Conclusions}
\label{sec:conclusions}
In this work, we focused on symmetric teleparallel gravity, a modification where non-metricity describes gravity while curvature and torsion vanish. By using the coincident gauge, a system of coordinates for which the affine connection vanishes identically, we derived the most general components of the connection for a stationary and spherically symmetric spacetime. These components coincide with the second set of \cite{DAmbrosio:2021zpm} while the first set arises as a special case of the second. The importance of our derivation lies on the fact that we obtained this result by direct integration from the coincident gauge and our connection covers both sets without taking any limits.

We then considered an interesting class of symmetric teleparallel theories, the Newer GR where the non-metricity scalar is decomposed into its irreducible components and five parameters are introduced. We discussed that in order for the theory to be in agreement with the PPN parameters, there are two families of constraints \cite{Flathmann:2020zyj}. The first one corresponds to the Type 1 theory while the second one in combination with the ghost-free condition \cite{Bello-Morales:2024vqk} corresponds to the Type 2 theory. Both theories have only one free parameter. For both types of Newer GR, by comparing  with the literature \cite{Hohmann:2018wxu}, we concluded that gravitational waves have two polarization modes, the same as in GR. We derived the field equations and we investigated solutions for a spherically symmetric ansatz. By constraining some of the free functions and parameters, we were able to find two interesting families of solutions which we studied.

The first one is a solution of Type 1 Newer GR while the second one is a common solution of both Type 1 and Type 2 Newer GR. Both of them correspond to exotic objects that do not have an analogue in GR.  For the first solution, we found that its Komar mass is zero and thus, an observer at infinity does not realize the existence of the object or perceives it as massless. Then, we specified one more free function in order to obtain asymptotic flatness and we found one spacetime singularity whose location depends on a constant. For the second solution, we found that its Komar mass is infinite, it is not asymptotically flat and there is one spacetime singularity.  For both solutions, we computed the gravitational redshift and we concluded that they admit no horizons. We also discussed how a particle moves near these objects. We introduced the constants of motion and an effective potential. In the case of particles moving in the radial direction, we computed the expansion of the beam and we derived the Raychaudhuri equations. For both solutions, the expansion is positive for outward motion and negative for inward motion and the worldlines converge as time passes. These conclusions are in agreement with General Relativity, the focusing theorem and the attractiveness of gravity. Regarding bound orbits, we did not find any for either of the solutions. We also found that the light deflection for the first solution vanishes at large distances as for a flat spacetime and that for the second solution there is no deflection because the light ray hits the singularity. Finally, we constructed the Penrose diagrams for these spacetimes.

As a future work, it would be challenging to review the field equations in Newer GR and possibly find more solutions or even attempt to solve them generally. In this case, it would be interesting to realize if this theory predicts objects already introduced in GR such as black holes or wormholes, so we are able to compare these solutions with their GR counterparts.

\begin{acknowledgments}
The authors gratefully acknowledge the full financial support by the Estonian Research Council through the Center of Excellence TK202 ``Fundamental Universe''.
\end{acknowledgments}

\appendix
\section{Field equations in Type 1 Newer GR}
\label{appendix1}
The following equations are the non-zero components of the field equations in Type 1 Newer GR for the spherically symmetric ansatz:
\begin{eqnarray}
E_{tt}&=&\frac{1}{4 r^{4} B^{3} (m  f_{1}-l  f_{2})^{2}}(-4 r^{3} A^{2} B' (r (2 V-1) f_{1}+2) (m  f_{1}-l  f_{2})^{2} \nonumber \\
&&+4 m  r^{2} (2 V-1) A^{2} B^{2} B' (m  f_{1}-l  f_{2})+r^{2} A^{2} B (2 f_{1} f_{2} ((2 V-1) ((l ^{2} r^{2}-3 A^{2})f'_{2} \nonumber \\
&&+f_{2}^{2} (l ^{2} r^{2}-3 A^{2})-3 l  m  r^{2} f'_{1}+2 l  m  r f_{2})-4 l  m)\nonumber \\
&&+f_{1}^{2} ((2 V-1) (f_{2}^{2} (6 A^{2}+l  r^{2} (l -4 m ))+2 m  r^{2} (2 m  f'_{1}-l  f'_{2})-4 m  r (l +m ) f_{2})+4 m ^{2})\nonumber \\
&&+2 f_{2}^{2} ((2 V-1) (3 A^{2}+l ^{2} r^{2})f'_{1}+2 l ^{2})+m ^{2} r^{2} (2 V-1) f_{1}^{4}+2 m  r (2 V-1) f_{1}^{3} (2 m +r (m -l ) f_{2})) \nonumber \\
&&+m ^{2} (2 V-1) B^{5} (A^{2}+2 l  r^{2} (m -l ))+B^{3} (-4 m  r^{2} (2 V-1) A^{2} (m  f'_{1}-l  f'_{2})+\nonumber \\
&&2 m  r f_{1} (A^{2} (2 m +r f_{2} (4 l -m +2 m  V)-4 m  V)-l  r^{2} (2 V-1) (2 m  (m -l )+l  r (l +2 m ) f_{2}))\nonumber \\
&&+m ^{2} r^{2} f_{1}^{2} (l  r^{2} (2 V-1) (l +2 m )-2 (2 V+1) A^{2})-4 l  m  r (2 V-1) f_{2} (l  r^{2} (l -m )-A^{2})\nonumber \\
&&+f_{2}^{2} (-2 l  r^{2} A^{2} (l +m +2 l  V-2 m  V)+(3-6 V) A^{4}+l ^{3} r^{4} (2 V-1) (l +2 m )))) \\ \nonumber \\
E_{tr}&=&\frac{1}{2 r^{4} A B (m  f_{1}-l  f_{2})^{2}}(2 V-1) (-l  r^{4} B A' (m  f_{1}-l  f_{2})^{2}+3 r^{2} A^{2} B f_{2} A' (m  f_{1}-l  f_{2})\nonumber \\
&&+r^{2} A (-l  r^{2} B' (m  f_{1}-l  f_{2})^{2}+r B (m  f_{1}-l  f_{2}) (-l  (-m  r f'_{1}+l  r f'_{2}+2 (m -l ) f_{2}+l  r f_{2}^{2})\nonumber \\
&&+m  r (l +m ) f_{1}^{2} -f_{1} (2 m  (m -l )+l ^{2} r f_{2}))+m  B^{3} (m  (m -2 l ) f_{1}+l  (3 m -2 l ) f_{2}))\nonumber \\
&&-A^{3} f_{2} r^{2} B' (m  f_{1}-l  f_{2})+r^{2} B (-m  f'_{1}-2 m  f_{1}^{2}+(3 m -2 l ) f_{1} f_{2}+l (f'_{2}+f_{2}^{2}))+m  B^{3})) \\ \nonumber \\
E_{rr}&=&\frac{1}{4 r^{4} A^{2} (m  f_{1}-l  f_{2})^{2}}B^{2} ((2 m  r f_{1} (2 m  r (2 V-1) A A'+A^{2} (r f_{2} (-4 l +m -2 m  V) \nonumber \\
&&+2 m  (2 V-1))-l  r^{2} (2 V-1) (2 m  (m -l )+l  r (l +2 m ) f_{2}))+f_{2} (4 l  m  r^{2} (1-2 V) A A' \nonumber \\
&&+2 l  r A^{2} (2 m +r f_{2} (l +m +2 l  V-2 m  V)-4 m  V)+(2 V-1) A^{4} f_{2}+l ^{2} r^{3} (2 V-1) (4 m  (m -l )\nonumber \\
&&+l  r (l +2 m ) f_{2}))+m ^{2} r^{2} f_{1}^{2} ((4 V+2) A^{2}+l  r^{2} (2 V-1) (l +2 m ))) \nonumber \\
&&+r^{2} A (-4 r A' (r (2 V-1) f_{1}+2) (m  f_{1}-l  f_{2})^{2}+A (m  f_{1}-l  f_{2}) (2 l  f_{2} (r^{2} (2 V-1) f'_{1}+2)\nonumber \\
&&+m  r^{2} (2 V-1) f_{1}^{3}-2 f_{1} (2 m +r (2 V-1) (l  r f'_{2}+f_{2} (-4 l +2 m +l  r f_{2})))+r (2 V-1) f_{1}^{2} (-4 m -r (l -2 m ) f_{2}))\nonumber \\
&&+2 (2 V-1) A^{3} f_{2} (f_{2} f'_{1}-f_{1} (f'_{2}+f_{2}^{2})+f_{1}^{2} f_{2})) -3 m ^{2} (2 V-1) B^{4} (A^{2}+2 l  r^{2} (l -m ))) \\ \nonumber \\
E_{\vartheta\vartheta}&=&\frac{1}{4 r^{2} A^{3} B^{3} (m  f_{1}-l  f_{2})^{2}}(-2 l  m  r^{4} (2 V-1) B^{3} (l -m ) A' (l  f_{2}-m  f_{1})\nonumber \\
&&+2 r^{2} A^{2} (m  f_{1}-l  f_{2}) (2 r^{2} A' B' (m  f_{1}-l  f_{2})+m  (1-2 V) B^{3} A'+r B (2 r A'' (l  f_{2}-m  f_{1})\nonumber \\
&&+A' (m  f_{1} (r (2 V-1) (f_{1}-f_{2})-2)+2 l  f_{2})))+l  r^{4} (2 V-1) A B^{2} (2 m  (l -m ) B' (l  f_{2}-m  f_{1})\nonumber \\
&&+B (-2 m  (l -m ) (l  f'_{2}-m  f'_{1})+m ^{2} (l +2 m ) f_{1}^{2}-2 l  m  (l +2 m ) f_{1} f_{2}+l ^{2} (l +2 m ) f_{2}^{2})) \nonumber \\
&&+A^{3} (2 r^{3} B' (m  f_{1}-l  f_{2}) (f_{1} (2 m +r (2 V-1) (m  f_{1}+(m -2 l ) f_{2}))-2 l  f_{2})\nonumber \\
&&-2 m  r^{2} (2 V-1) B^{2} B' (m  f_{1}-l  f_{2})-r^{4} (2 V-1) B (f_{1}^{2} (2 m  (m  f'_{1}+(m -2 l ) f'_{2})+l  (l -4 m ) f_{2}^{2}) \nonumber \\
&&+2 l  f_{1} f_{2} (l  (f'_{2}+f_{2}^{2})-m  f'_{1})+2 l  (l -m ) f_{2}^{2} f'_{1}+m ^{2} f_{1}^{4}+2 m  (m -l ) f_{1}^{3} f_{2})\nonumber \\
&&+2 m  r^{2} (2 V-1) B^{3} (m  f'_{1}-l  f'_{2})+m ^{2} (2 V-1) B^{5})+(1-2 V) A^{5} B^{3} f_{2}^{2})
\end{eqnarray}

\section{Field equations in Type 2 Newer GR}
\label{appendix2}
The following equations are the non-zero components of the field equations in Type 2 Newer GR for the spherically symmetric ansatz:
\begin{eqnarray}
E_{tt}&=&-\frac{1}{2 B^{4}}(8VABA' B'+4 V B^{2} (A'^{2}+B^{2} (3 l +m )^{2})-\frac{8 V A B^{2} (r A''+2 A')}{r}\notag \\
&&+20 V A^{2} B'^{2}+\frac{1}{r (m  f_{1}-l  f_{2})}(2 A^{2} B (4 r V B'' (l  f_{2}-m  f_{1})+B' (8 r V (l  f_{2}'(r)-m  f_{1}'(r))\notag \\
&&+f_{1} (m -8 r V (m -3 l ) f_{2}+8 m  V)-24 m  r V f_{1}^{2}+8 l  r V f_{2}^{2}-f_{2} (l +8 l  V))))\notag \\
&&+\frac{1}{r^{2} (m  f_{1}-l  f_{2})^{2}}(A^{2} B^{2} (8 l  r^{2} V f_{2} (l  f_{2}''(r)-m  f_{1}''(r))+f_{1}^{2} (4 r^{2} V (2 m  (6 m  f_{1}'(r)+(m -3 l ) f_{2}'(r))\notag \\
&&+(9 l ^{2}-12 l  m +m ^{2}) f_{2}^{2})-m ^{2})-4 r^{2} V (m  f_{1}'(r)-l  f_{2}'(r))^{2}+l  f_{2}^{2} (8 r^{2} V ((3 l -m ) f_{1}'(r)+2 l  f_{2}'(r))-l )\notag \\
&&+2 f_{1} (4 m  r^{2} V (m  f_{1}''(r)-l  f_{2}''(r))+f_{2} (l  m +4 r^{2} V (m  (m -9 l ) f_{1}'(r)+3 l  (l -m ) f_{2}'(r)))+4 l  r^{2} V (3 l -m ) f_{2}^{3}) \notag \\
&&+24 m  r^{2} V (m -3 l ) f_{1}^{3} f_{2}+36 m ^{2} r^{2} V f_{1}^{4}+4 l ^{2} r^{2} V f_{2}^{4}))+\frac{A^{2} B^{4}}{r^{2}}) \\ \notag \\
E_{tr}&=&\frac{1}{r A B (m  f_{1}(r)-l  f_{2})}4 V (3 l +m ) (r B A' (m  f_{1}-l  f_{2})+A (r B' (m  f_{1}-l  f_{2})+B (-m  r f'_{1}\notag \\
&&-3 m  r f_{1}^{2}+2 m  f_{1}+r (3 l -m ) f_{1} f_{2}+l  r (f'_{2}+f_{2}^{2})-2 l  f_{2})))+B (-m  r f_{1}-3 m  r f_{1}^{2}+2 m  f_{1}+r (3 l -m ) f_{1} f_{2}\notag \\
&&+l  r (f_{2}+f_{2}^{2})-2 l  f_{2})))) \\ \notag \\
E_{rr}&=&-\frac{4 V A' B'}{A B}-\frac{2 V (A'^{2}+B^{2} (3 l +m )^{2})}{A^{2}}+\frac{1}{r A (m  f_{1}-l  f_{2})}(4 r V A'' (l  f_{2}-m  f_{1})\notag \\
&&+A' (-8 V (-m  r f'_{1}+l  r f'_{2}+l  r f_{2}^{2}-3 l  f_{2})+f_{1} (8 r V (m -3 l ) f_{2}-m  (24 V+1))+24 m  r V f_{1}^{2}+l  f_{2}))\notag \\
&&+\frac{6 V B'^{2}}{B^{2}}-\frac{4 V (r B''+2 B')}{r B}+\frac{B^{2}}{2 r^{2}}+\frac{1}{2 r^{2} (m  f_{1}-l  f_{2})^{2}}(-12 r^{2} V (m  f'_{1}-l  f'_{2})^{2}+l  f_{2}^{2} (8 r^{2} V (3 l +m ) f'_{1}\notag \\
&&-l  (32 V+1))-2 f_{1} (f_{2} (4 r^{2} V (3 l +m ) (m  f'_{1}+l  f'_{2})-l  m  (32 V+1))+4 l  r^{2} V (3 l -m ) f_{2}^{3}\notag \\
&&+4 m  r V (-m  r f''_{1}-4 m  f'_{1}+l  r f''_{2}+4 l  f'_{2})+16 l  r V (2 m -3 l ) f_{2}^{2})-36 m ^{2} r^{2} V f_{1}^{4}-4 l ^{2} r^{2} V f_{2}^{4}\notag \\
&&+8 l  r V f_{2} (-m  r f''_{1}-4 m  f'_{1}+l  r f''_{2}+4 l  f'_{2})+f_{1}^{2} (4 r V (2 m  r (3 l +m ) f'_{2}-r (9 l ^{2}-12 l  m +m ^{2}) f_{2}^{2}+8 m  (m -6 l ) f_{2})\notag \\
&&-m ^{2} (32 V+1))+24 m  r V f_{1}^{3} (4 m +r (3 l -m ) f_{2})+32 l ^{2} r V f_{2}^{3}) \\ \notag \\
E_{\vartheta\vartheta}&=&\frac{1}{2} r (\frac{r (8 V+1) A' B'}{A B^{3}}+\frac{4 r V A'^{2}}{A^{2} B^{2}}-\frac{r (8 V+1) A''+(16 V+1) A'}{A B^{2}}-\frac{4 r V (3 l +m )^{2}}{A^{2}}+\frac{20 r V B'^{2}}{B^{4}}\notag \\
&&+\frac{1}{B^{3} (m  f_{1}-l  f_{2})}(8 r V B'' (l  f_{2}-m  f_{1})+B' (16 V (-m  r f'_{1}+l  r f'_{2}+l  r f_{2}^{2}-l  f_{2})+f_{1} (m -16 r V (m -3 l ) f_{2}\notag \\
&&+16 m  V)-48 m  r V f_{1}^{2}-l  f_{2}))+\frac{1}{B^{2} (m  f_{1}-l  f_{2})^{2}}(4 r V (2 l  f_{2} (l  f''_{2}-m  f''_{1})+f_{1}^{2} (2 m  (6 m  f'_{1}+(m -3 l ) f'_{2})\notag \\
&&+(9 l ^{2}-12 l  m +m ^{2}) f_{2}^{2})-(m  f'_{1}-l  f'_{2})^{2}+2 l  f_{2}^{2} ((3 l -m ) f'_{1}+2 l  f'_{2})+2 f_{1} (m  (m  f''_{1}-l  f''_{2})+f_{2} (m  (m -9 l ) f'_{1}\notag \\
&&+3 l  (l -m ) f'_{2})+l  (3 l -m ) f_{2}^{3})+9 m ^{2} f_{1}^{4}+6 m  (m -3 l ) f_{1}^{3} f_{2}+l ^{2} f_{2}^{4})))
\end{eqnarray}

\bibliography{references}

\end{document}